\documentclass{aa}
\usepackage{natbib}
\bibpunct{(}{)}{;}{a}{}{,} 
\usepackage{hyperref}
\usepackage{aalongtable}
\usepackage{graphicx}
\usepackage{txfonts}
\usepackage{balance}
\usepackage{wrapfig}
\usepackage{latexsym}
\newcommand{\nc}{\newcommand}  
\newcommand{\dn}{$\Delta_{\rm NLTE}$\,}
\nc{\teff}{$T_{\rm eff}$\,}  
\nc{\teffns}{$T_{\rm eff}$}  
\nc{\glog}{log\,$g$\,}  
\nc{\kms}{\,${\rm km\,s}^{-1}$\,}  
\nc{\mic}{$\xi_{\rm t}$\,}
\nc{\nod}{--}

\begin{document}

\title{{First stars XIII. Two extremely metal-poor RR Lyrae stars}
\thanks {Based on observations made with the ESO Very Large Telescope 
at Paranal Observatory, Chile (Large Programme ``First Stars'', 
ID 165.N-0276(A); P.I. R. Cayrel).}
}

\author{
C. J. Hansen\inst {1,10}\and
B.~Nordstr\"om \inst {1} \and 
P. Bonifacio\inst {2,3,4} \and
M. Spite\inst {2} \and
J.~Andersen\inst {1,7} \and
T.~C.~Beers\inst {5} \and
R. Cayrel\inst {2} \and
F. Spite \inst {2} \and
P. Molaro\inst {3,4} \and
B.~Barbuy\inst {8} \and
E. Depagne \inst{9} \and
P. Fran\c cois  \inst{2} 
V. Hill\inst {11} \and 
B. Plez\inst {5} \and
T. Sivarani \inst{5} 
} 

\institute{ 
         The Niels Bohr Institute, Astronomy, Juliane Maries Vej 30,
         DK-2100 Copenhagen, Denmark\\
   \email{camjh@astro.ku.dk, birgitta@astro.ku.dk,ja@astro.ku.dk}
  \and
Observatoire de Paris, GEPI,
             F-92195 Meudon Cedex, France\\
\email{Piercarlo.Bonifacio@obspm.fr, Roger.Cayrel@obspm.fr}\\
   \email{Monique.Spite@obspm.fr, Francois.Spite@obspm.fr}\\
   \email{Patrick.Francois@obspm.fr }
         \and
CIFIST Marie Curie Excellence Team
\and
    Istituto Nazionale di Astrofisica - Osservatorio Astronomico di
Trieste,
    Via Tiepolo 11, I-34131
             Trieste, Italy\\
   \email{molaro@ts.astro.it}
\and
 Department of Physics \& Astronomy and JINA: Joint Institute for Nuclear
Astrophysics, Michigan State University,
             East Lansing, MI 48824, USA\\
   \email{thirupati@pa.msu.edu,beers@pa.msu.edu}
         \and
GRAAL, Universit\'e de Montpellier II, F-34095 
Montpellier
             Cedex 05, France\\
   \email{Bertrand.Plez@graal.univ-montp2.fr}
         \and
     Nordic Optical Telescope, Apartado 474, ES-38700 Santa Cruz de 
     La Palma, Spain\\
   \email{ja@not.iac.es}
         \and
    Universidade de S\~ao Paulo, Departamento de Astronomia,
Rua do Mat\~ao 1226, BR-05508-900 S\~ao Paulo, Brazil\\
   \email{barbuy@astro.iag.usp.br}
               \and
Las Cumbres Observatory, Goleta, CA 93117 USA \\
    \email{edepagne@lcogt.net}
       \and 
         European Southern Observatory (ESO),
         Karl-Schwarschild-Str. 2, D-85748 Garching b. M\"unchen, 
Germany\\
   \email{cjhansen@eso.org}
        \and          	
     CASSIOP\'EE, Universit\'e de Nice Sophia Antipolis, Observatoire de la C\^ote d'Azur, BP 4229, F-06304 Nice Cedex 4, 
France\\
    \email{Vanessa.Hill@oca.eu}    
}
\offprints{C. J. Hansen}
\date{Received xxx; Accepted xxx}
\abstract{
The chemical composition of extremely metal-poor stars (EMP stars; [Fe/H]$<$ 
$\sim-3$) is a unique tracer of early nucleosynthesis in the Galaxy. As such stars are rare, we wish to find classes of luminous stars which can be studied at high spectral resolution.
}
{
We aim to determine the detailed chemical composition of the two EMP stars CS~30317-056 and CS~22881-039, originally thought to be red horizontal-branch (RHB) stars, and compare it to earlier results for EMP stars as well as to nucleosynthesis yields from various supernova (SN) models. In the analysis, we discovered that our targets are in fact the two most metal-poor RR Lyrae stars known.}
{Our detailed abundance analysis, taking into account the variability of the stars, is based on VLT/UVES spectra ($R \simeq 43,000$) and 1D LTE OSMARCS model atmospheres and synthetic spectra. For comparison with SN models we also estimate NLTE corrections for a number of elements.}
{We derive LTE abundances for the 16 elements O, Na, Mg, Al, Si, S, Ca, 
Sc, Ti, Cr, Mn, Fe, Co, Ni, Sr and Ba, in good agreement with earlier 
values for EMP dwarf, giant and RHB stars. Li and C are not detected in either star. 
NLTE abundance corrections are newly calculated for O and Mg and taken from 
the literature for other elements. The resulting abundance pattern is best matched by model yields for supernova explosions with high energy and/or significant asphericity effects.} 
{Our results indicate that, except for Li and C, the surface composition
of EMP RR Lyr stars is not significantly affected by mass loss, mixing or diffusion processes; hence, EMP RR Lyr stars should also be useful tracers 
of the chemical evolution of the early Galactic halo. The observed abundance 
ratios indicate that these stars were born from an ISM polluted by energetic,
massive ($25-40{\rm M}_{\odot}$) and /or aspherical supernovae, but the 
NLTE corrections for Sc and certain other elements do play a role in the 
choice of model.}
\maketitle

\titlerunning{First stars XIII. Two EMP RR Lyrae stars}
\authorrunning{C.J. Hansen et al.}
%
\section{Introduction}

The most direct information on the nature of early star formation and nucleosynthesis 
in the Galaxy is provided by the chemical composition of very metal-poor stars
\citep{firstV, firstVIII, lai08, Bonifaciodwarf}. Much of this information
derives from studies of extremely metal-poor (EMP) giant stars, which are bright and display spectral lines of many elements.  However, the surface chemical composition of giants can be affected by products of nuclear burning in their 
interiors, which have been mixed to the surface at a later stage \citep{firstVI, first9}.

Alternative tracers exist, but have not been studied in the same detail as the giants. EMP turnoff stars are not susceptible to surface mixing, but are intrinsically fainter, and the hotter stars may be affected by atomic diffusion, levitation and gravitational settling. In addition, many lines of heavy 
elements are not visible in spectra of EMP
turnoff stars due to the higher opacity of their atmospheres. 
Nevertheless, the abundances observed
in giants and dwarfs of similar metallicity generally agree well, although some
discrepancies are seen. These are primarily ascribed to 3D and NLTE effects -- see e.g., \citet{Bonifaciodwarf} and \citet{andrMg10}. 

Our original aim was to test whether red horizontal-branch (RHB) stars could be used as tracers of early Galactic nucleosynthesis. To this end, we performed a detailed analysis of the two EMP stars CS~22881-039 and
CS~30317-056, originally discovered in the HK Survey of Beers et al.
\citep{Beers85,Beers1992} and classified as RHB stars from their colours. 
During the analysis we discovered the stars to be variable in radial velocity, and subsequently that although the effective temperatures of both stars are clearly outside the fundamental red edge of the instability strip defined by \citet{preston}, both are in fact RR Lyrae variables.
CS~22881-039 was studied already by \citet{preston}, but from limited spectroscopic data that did not enable them to discover its variability. 

The paper is organised as follows: Section 2 describes our spectroscopic 
observations and Section 3 their relation to the pulsational phase of the stars at the times of observation. On this basis, Section 4 describes our derivation of stellar parameters, while Section 5 discusses the LTE
abundance analysis and summarises our adopted NLTE abundance corrections. 
Section 6 compares our abundances with earlier values for giants,
dwarfs, and RHB stars, and Section 7 compares them with nucleosynthesis yields from recent SN models. Sections 8 and 9 present our discussion and conclusions.


\section{Observations and Data Reduction}

Our two targets were observed as part of our ESO Large Programme 'First
Stars', using the VLT Kueyen telescope and UVES spectrograph at a resolving
power $R \sim 43,000$, similar to that used by \citet{preston}. The wavelength range 334-1000 nm is covered almost completely, using the blue and red cameras of UVES simultaneously. The observations and data
reduction were performed exactly as in the rest of this
programme \citep[see][for descriptions]{firstI,firstV}, making our results
directly comparable with those published previously.


\begin{table*}[th]
\caption{Log of the observations. The first and last spectrum of CS30317-056 were used only to determine the RV.}
\label{logobs}
\begin{center}
\begin{tabular}{cccccccc}
\hline\hline
Spectrum & $V$ & $\lambda$ Setting [nm] & Date & MJD  & Exp.time [s] & Phase &        V$_{Barycentric}$ [km/s] \\
\hline
CS22881-39 & 15.1 &&&&&& \\
\hline                   
152-10\_B  &&  396 &   1/6/2001  &   52061.3657  &   4800 & 0.86 & +90.0 \\
152-14\_V  &&  573 & \\  
\hline                          
CS30317-056 & 14.0 &&&&&& \\                            
\hline
222-6\_V && 573 &    9/8/2000 &   51765.9703  &  3600 &   0.13  &  -62.9 \\
151-7\_B && 396 &   31/5/2001 &   52060.9970  &  3600 &   0.28  &  -53.0 \\
151-8\_V && 573 &\\
152-4\_B && 396 &    1/6/2001 &   52061.0864  &  3600 &   0.40  &  -45.3 \\
152-8\_R && 850 &\\
152-5\_B && 396 &    1/6/2001 &   52061.1323  &  3600 &  0.47 &   -41.6 \\
152-9\_R && 850 &\\
153-5\_B && 396 &    2/6/2001 &   52062.2195  &  600  &  0.92  &  -51.4 \\
153-5\_V && 573 &\\
\hline\hline  
\end{tabular}
\end{center}
\end{table*}

Four spectra of CS~30317-056 were obtained in May/June 2001. The last of these has very low S/N and was used only to determine the radial velocity, as was a spectrum obtained in August 2000 with the red camera, covering only the wavelength range 460 -- 670nm. For CS~22881-039, only a single spectrum from June 2001 is available. 

\begin{figure}[ht]
\begin{center}
\includegraphics[width=0.4\paperwidth]{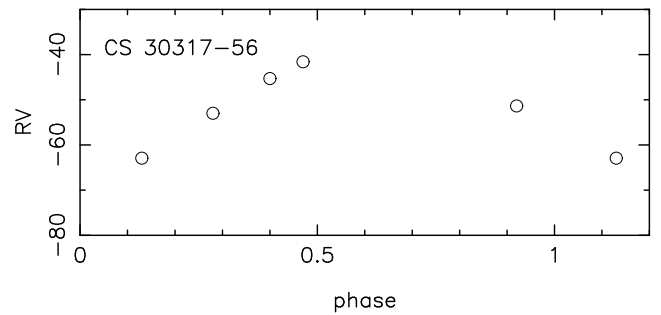}
\caption{The radial-velocity curve of CS30317-56, counting phase from maximum light.}
\label{RV}
\end{center}
\end{figure}

\section{CS~22881-039 and CS30317-056 as RR Lyr stars}

Following our discovery of the radial velocity variability of our two stars, they were also identified as photometric variables in the database of the WASP project (A. Collier Cameron, priv. comm.). The periods of CS~22881-039 are $P=0.66876$ and $P=0.74851$ days, typical of RRab variables as expected 
for ``red'' RR Lyr stars \citep[e.g.][p. 45]{Smi95}. 
The amplitude of CS~22881-039 is $\sim 0.8$ mag, typical of an RRb variable. For CS30317-056 the amplitude is only $\sim 0.35$ mag, small for an RRb and more like an RRc; however, the overtone RRc pulsators have less asymmetric light curves with periods $P < 0.5$ days and are generally hotter than CS30317-056. Several RRb stars with small amplitude have been observed by \citep{Nem04} in the metal-poor globular cluster NGC~5053, and the metallicity of our stars is substantially lower than that of any known globular cluster.

The WASP data enable us to compute the phase $\phi$ of our UVES observations, counting phase from maximum light. Both light curves then rise steeply from phase $\phi \sim 0.8$ to maximum, with a broad minimum in the phase interval $\phi \sim 0.3-0.8$. Fortunately our spectra of both stars were taken near minimum light and thus in a relatively static phase, favourable for a spectroscopic analysis \citep{kolenberg}; the two observations near $\phi =$ 0.1 and 0.9 in Fig. \ref{RV} were not used in the abundance analysis.

Table \ref{logobs} lists the times of the observations from which we derived radial velocities, and Fig. \ref{RV} shows the radial velocity curve of CS~30317-56.

Only few RR Lyr stars have high-dispersion spectroscopic analyses, although they play an important role in the determination of distances and their absolute magnitude depends on metallicity -- see \citet{LHL96}, \citet{SPS01}, \citet{preston}, and \cite{PAP09}. In fact, CS~22881-039 and CS~30317-056 seem to be the most metal-poor RR Lyr stars analysed in any detail so far. Thus, their properties are important for our understanding of the physics of RR Lyr stars and the location of the instability strip in the HR diagram as a function of metallicity.


\begin{table*}[ht]
\begin{center}
\caption{ Atmospheric parameters for the target stars. Disregarding 15 nearly 
saturated Fe~I lines in CS 22881-039 (22\% of all Fe~I lines) reduces the scatter in [Fe/H] by 37\% (numbers in parenthesis).}
\label{stellartab}
\centering
\begin{tabular}{cccccccccccc}
\hline
\hline
Star & \teff & $\sigma$ & \glog & $\sigma$ &  \mic & $\sigma$ & [Fe/H] & 
$\sigma$ & No. of lines & Mass & $\sigma$\\
     & K  & K  & cgs  & cgs & \kms & \kms  &  & dex & (Fe~I, Fe~II) & 
$M_{\odot}$ & $M_{\odot}$  \\
\hline
 \object{ CS~30317-056  } &  6000 & 200& 2.00 & 0.25 & 3.0 & 0.1 & -2.85 &  0.15 & 113, 4 & 0.7 & 0.1\\
 \object{ CS~22881-039  } & 5950 & 150 & 2.10 & 0.3 & 3.0 & 0.1 & -2.75 & 0.3(0.19) & 68(53), 6 & 0.57 & 0.12\\
\hline
\hline
\end{tabular}
\end{center}
\end{table*}

\section{Stellar Parameters\label{atm}}

RR Lyr stars are pulsating variables with large variations of \teff and \glog with pulsation phase: Following \citet{PAP09}, the change in temperature can reach 2000~K; in \glog up to 1.5 dex. It is thus very important to determine the parameters of the atmosphere of the stars {\it at the exact moment} of the observations. 

A rich Fe~I spectrum is a rather good thermometer: The derived abundance must be independent of the excitation potential of the line used. In this paper we have determined \teff for each spectrum from the LTE excitation equilibrium of Fe~I; the estimated uncertainty is about 200~K. Only the blue spectra were used in the determination, and only lines with an equivalent width larger than 1~pm. 

\begin{figure}[ht]
\begin{center}
\includegraphics[width=0.4\paperwidth]{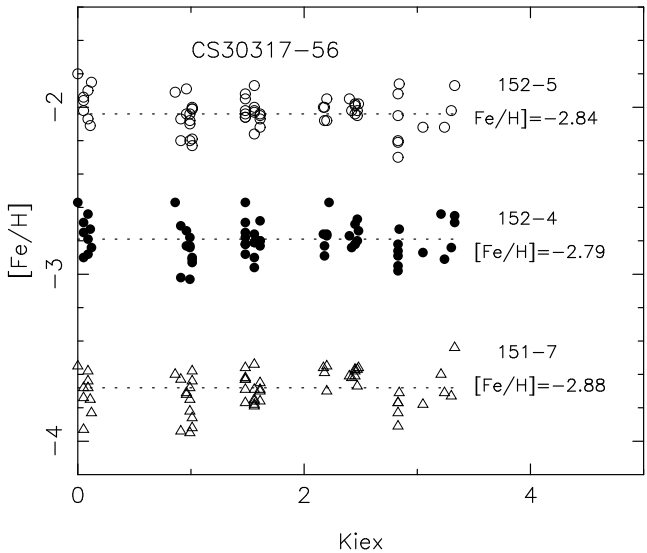}
\caption{[Fe/H] vs. $\rm \chi_{ex}$ for our three blue spectra of CS~30317-56 (152-5, 152-4 and 151-7). For clarity, the plots for 151-7 and 152-5 have been shifted down and up by 0.8 dex, respectively. The computations were with the same OSMARCS model: \teff=6000~K,  \glog =2.0, [M/H]=--2.85 dex, $\rm\xi_{t}=3 ~km s^{-1}$. The mean abundance from the Fe~I lines is given to the right of each plot. Abundances from Fe~II are $\rm [Fe/H]_{151-7}=-2.84$ dex, $\rm [Fe/H]_{152-4}=-2.84$ dex, and $\rm [Fe/H]_{152-5}=-2.90$ dex. Accordingly, within the errors, the same model can be adopted for all spectra.
}
\label{fekiex}
\end{center}
\end{figure}

Fig. \ref{fekiex} presents the Fe~I abundance from the three blue spectra of  CS~30317-056 as a function of excitation potential ($\chi_{ex}$), computed with the same model: \teff= 6000~K, \glog =2.0, [M/H]=--2.85 dex and microturbulent velocity, $\rm\xi_{t}=3 ~km s^{-1}$. 
As seen, for a common temperature of 6000~K there is no trend of Fe I abundance with the excitation potential of the lines for all the spectra.  The star was thus rather stable during the observations, as expected for observations close to light minimum.

CS~22881-039 was observed only once, just before the steep rise in luminosity (at phase 0.86). The effective temperature derived from our spectrum is also rather low for an RR Lyr star, \teff= 5950~K. The barycentric radial velocity is $\sim$ 90.0 km/s. 

OSMARCS LTE model atmospheres \citep{OSMARCS,Gus08} and the 1D LTE synthetic 
spectrum code {\tt turbospectrum} \citep{turbospectrum} were used to derive stellar parameters and abundances. The procedure is described in \citet{firstV}. 

The microturbulent velocity was determined by requiring that the derived Fe~I
abundance be independent of the equivalent width of the lines. 

Surface gravities (\glog) were obtained from the ionization balance of Fe (as in \citet{preston}). 
With \teff=6000 K and \glog=2.0 there is good agreement between the abundances derived from Fe I and Fe II lines in all three spectra (Table \ref{abfe}; $\rm \Delta\Sigma = [Fe_{II}/H] - [Fe_{I}/H] ~ < 0.1 dex$).


\begin{table}[htb] 
\caption{[Fe/H] as computed from Fe~I and Fe~II lines in our three ``blue'' spectra of CS~30317-56, for \teff=6000~K and \glog=2.0. n1 and n2 are the number of lines used in each determination. The phase is given for each spectrum.}  
\label{abfe} 
\begin{center}
\begin{tabular}{l c c c c c c c c }
\hline\hline
Spectrum &  $\rm[Fe_{I}/H]$ &n1 & $\rm[Fe_{II}/H]$  & n2&$\Delta \Sigma$ & Phase \\
\hline
151-7&    -2.89 & 52&  -2.84& 9  &    +0.05  & 0.28  \\
152-4&    -2.80 & 55&  -2.84& 8  &    -0.04  & 0.40  \\
152-5&    -2.82 & 54&  -2.90& 7  &    -0.08  & 0.47 \\
\hline\hline   
\end{tabular}  
\end{center}  
\end{table}   

In pulsating stars such as Cepheids or RR Lyr stars, the Baade-Wesselink method can also be used to determine the luminosity and thus the surface gravity of the star. However, for RR Lyraes it has been shown that precise infrared light curves, which are not available for our stars, must be used (e.g. \citet{Smi95} p. 31, \citet{CCF1992}).	

The final atmospheric parameters for each star are given in Table \ref{stellartab}; [Fe/H] is the mean value over all measured Fe~I and Fe~II lines.  

Non-LTE effects have been observed in RR Lyr stars. If the luminosity is known (e.g. from a Baade-Wesselink study), the gravity can be derived directly. The mean abundance of Fe I is then found to be lower than that derived from the Fe~II lines (overionisation of iron). Following \citet{LHL96}, an abundance discrepancy of $\Delta \Sigma \approx -0.2$ is seen in RR Lyr stars when the gravity is fixed from the luminosity; reducing of $\Delta \Sigma$ to 0.0 requires a change of \glog by about 0.6~dex.  Thus, the true gravities (corresponding to the luminosity of the stars) are probably about 0.6~dex larger than the value given in Table \ref{stellartab}.

The masses given in Table \ref{stellartab} were estimated from the
evolutionary tracks for [M/H]= $-2.27$ by \citet{cassisi}. 
As discussed
by \citet{preston}, the changes in \glog derived from these tracks are comparable to the observational uncertainty over a reasonable range in [Fe/H],
and we estimate the uncertainty of the derived masses to be $\sim\pm$0.1 M$_{\odot}$.

\begin{figure}[h]
\begin{center}
\includegraphics[width=0.45\textwidth]{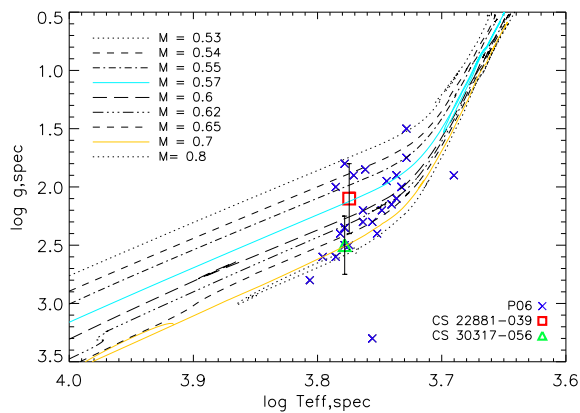}
\caption{Spectroscopically determined log T$_{\rm{eff}}$ vs. \glog and evolutionary tracks 
 for model masses $0.53 \leq \rm{M/M_{\odot}} \leq 0.8$ and [M/H] $= -2.27$
by \citet{cassisi}. CS~30317-056 is shown as a green triangle, CS~22881-039 as a
red square, and the RHB stars of \citet{preston} as blue $\times$-es.}
\label{prestonmasse}
\end{center}
\end{figure}

Figure \ref{prestonmasse} shows the \citet{cassisi} evolutionary tracks in the
log \teff -- \glog plane together with our spectroscopic values 
for CS~30317-056 and CS~22881-039. The \citet{preston} RHB stars are shown
for comparison; as seen, CS~30317-056 and CS~22881-039 fall well within the
range covered by these stars (i.e. outside the instability strip ending at the fundamental red edge at log T$_{eff} = 3.80$).

\section{Abundance Analysis}


\subsection{LTE abundances}

The LTE abundances were generally obtained from equivalent widths (listed in online Table \ref{linelist}), determined by fitting Gaussian profiles to observed lines 
from the lists of \citet{firstI}. For this, we used the genetic algorithm
{\tt fitline} of \citet{firstIII}. Lines were disregarded if they were blended
or very weak ($< 0.5$ pm); in case of doubt, the \citet{cayrel1988} formula 
was used to assess the uncertainty of the equivalent width.

If the Gaussian fit seemed uncertain, the equivalent width was remeasured by
direct integration in IRAF and checked with an LTE synthetic spectrum 
computed with {\tt turbospectrum} \citep{turbospectrum}. If the two derived
abundances differed significantly, the {\tt turbospectrum} result was adopted.
For a few elements, the abundance could only be determined from the
synthetic spectrum fit.

We did not detect the Li~I lines at 610.36 nm and 670.79 nm, even though the quality of the spectra was optimal in this region. The measured noise levels would allow to detect lines down to $\sim 4$ m\AA\ in equivalent width, which leads to the upper abundance limits for Li given in Table \ref{abuntab}.

We have tried to measure the C abundance of CS~22881-039 and CS~30317-056 from a fit to the G band of the CH molecule. Since we found that \teff and \glog did not change between the three blue spectra of CS~30317-056 (section \ref{atm}), we coadded the spectra (correcting for the radial velocity variation) to improve the S/N ratio in the G-band region.

\begin{figure}[ht]
\begin{center}
\includegraphics[width=0.4\paperwidth]{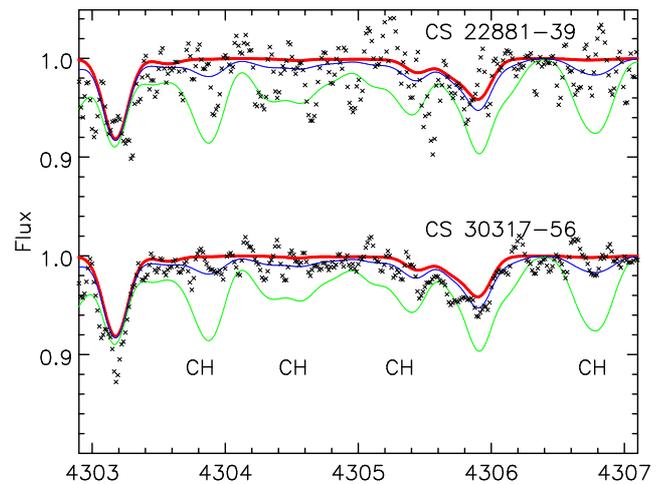}
\caption{Synthetic spectrum fits to the CH band in CS~22881-039 and 
CS~30317-056. Crosses: observations; thick line: spectra computed with no carbon; thin lines: spectra computed for $\rm log \epsilon(C)=6.5 $ and 7.2. The observed spectra are compatible with a complete absence of carbon; an upper limit to the abundance is $\rm log~ \epsilon(C) \lesssim 6.5 $.
}
\label{CHband}
\end{center}
\end{figure}

The LIFBASE program of Luque \& Crosley \citep{LC99} was used to compute line positions and {\sl gf}-values; excitation energies were taken from the line list of \citet{JLI96}. The spectral regions free from contamination by atomic lines are indicated in the figure. Carbon is not detected, neither in CS~22881-039 nor in CS~30317-056, but the noise is large. For both stars the upper limit to the C abundance is $\rm log~ \epsilon(C)=6.5 $ (Fig. \ref{CHband}) or $\rm [C/H] \lesssim -2$.

We have also tried to measure the Na abundance in CS~22881-39 and CS~30317-56. In CS~30317-56 we observe a strong variation of the shape of the Na D lines with pulsation phase, $\phi$: At $\phi \approx 0.30$ the D lines are relatively narrow and symmetric, while they are broad and asymmetric at $\phi = 0.13$.
Only the first red spectrum (taken together with one of the blue spectra used to determine \teff, log g, and $\rm \xi_{t}$) was applied to determine the Na abundance.

In CS 22881-39 the D lines are broad and asymmetric and unsuitable for an abundance determination. Since we have only one spectrum, we could not determine whether the asymmetry is due to shock waves in the atmosphere just before the steep rise to light maximum or to a blend by interstellar D lines.

The shape of the aluminium lines is very similar in the three blue spectra of CS~30317-56. Thus we have used the mean spectrum to fit the line profiles. (Since one of the Al lines is in the wing of the Balmer line $\rm H\epsilon$, it is better to determine the abundance from a synthetic spectrum fit).
We find [Al/Fe]=-0.77 for CS~22881-39 and [Al/Fe]=-0.51 for CS~30317-56. 

LTE abundances for the 16 elements  O, Na, Mg, Al, Si, S, Ca, Sc, Ti, Cr, Mn,
Fe, Co, Ni, Sr and Ba in CS~30317-056 and CS~22881-039 are listed in Table 
\ref{abuntab} together with the NLTE abundance corrections discussed in the 
following section. Errors of abundances based on only one line have been 
estimated from the uncertainty of the equivalent width. 
For an easier comparison with the other ``First Stars'' papers we have adopted the same (1D, LTE) solar reference abundances from \citet{GrevSauv}, which differ by typically only $\sim 0.1$ dex from the 3D+NLTE solar abundances by \citet{asp09}.


\begin{table*}
\caption{LTE abundance ratios and adopted NLTE corrections for our two RR Lyr stars. $\sigma$ is the standard deviation and N the number of individual lines. 'ul'  indicates an upper limit, '*' a value derived from a {\tt turbospectrum} fit, and a $\sim$ that the \dn \rm value is a rough estimate or only valid for some lines.}
\label{abuntab}
\addtolength\tabcolsep{-3pt}
\centering
\begin{tabular}{c c c c c c c c c}
\hline
\hline
Abundance & \object{ CS~22881-039  } &\begin{math} \sigma \end{math} & N & \dn & \object{ CS~30317-056  }& $ \sigma $& N \\
\hline
\ log $\epsilon$ (Li)& $\sim$ 0.6& ul &1&\nod&$\sim$ 1.0   &ul  & 1 \\
\ [C/Fe]      & 0.69  & ul & \nod   & \nod     &   0.79    & ul & \nod \\
\ [O/Fe]      & \nod  & \nod & \nod& -0.1      &   0.9    & 0.24 & 1  \\	      
\ [Na/Fe]     & -0.13 & 0.15 &  2  & -0.2/-0.1 &   \nod    & 0.13 & 2 \\
\ [Mg/Fe]     & 0.70  & 0.30 &  6  & $\sim$0.05& 0.48   & 0.15 & 6   \\	      
\ [Al/Fe]     & -0.74 & 0.05 &  2  & +0.7      & -0.66  & 0.15 & 2 \\
\ [Si/Fe]     & 0.01  & 0.05 &  1  & -0.05     & 0.05   & 0.05 & 1  \\    
\ [S/Fe]      & \nod  & \nod & \nod& -0.4      &  0.78    & 0.07 & 1 \\
\ [Ca/Fe]     & 0.29  & 0.30 &  5  & $\sim$0.19& 0.4    & 0.13 & 6 \\      
\ [ScII/Fe~II] & 0.11  & 0.06 &  3  & \nod      & 0.14   & 0.06 & 3\\
\ [TiI/Fe~I]   & 0.38  & 0.25 &  1  & \nod      & 0.63   & 0.16 & 6  \\ 
\ [TiII/Fe~II] & 0.35  & 0.25 &  17 & \nod      & 0.55   & 0.12 & 27 \\
\ [Cr/Fe]     & -0.32 & 0.12 &  4  & +0.4      & -0.18  & 0.05 & 5 \\     
\ [Mn/Fe]     & -0.59 & 0.05 &  2  & +0.6      & -0.63  & 0.05 & 3 \\
\ [Co/Fe]     & 0.14  & 0.14 &  2  &$\sim$+0.6 & 0.27   & 0.06 & 2 \\      
\ [Ni/Fe]     & -0.19 & 0.07 &  2  & \nod      & -0.13  & 0.05 & 3  \\
\ [SrII/Fe~II] & 0.00  & 0.15 &  1  & +0.6      & 0.03  & 0.10 & 2\\       
\ [BaII/Fe~II] & -0.62 & 0.09 &  1  &$\sim$+0.15& -0.32 & 0.05 & 2\\
\hline		   
\hline
\end{tabular}	    
\end{table*}

\subsection{NLTE abundance correction}\label{nlte}

NLTE effects become important when the aim is not just to compare results for different stars, but to confront the data with supernova models, which are independent of our incomplete understanding of stellar atmospheres.

NLTE calculations in the literature generally apply to either dwarfs or giants. We have therefore calculated NLTE abundance corrections 
\dn\footnote{\dn = $\log \epsilon_{\rm NLTE} - \log \epsilon_{\rm LTE}$.} 
for O and Mg for our stars and estimated other corrections from 
studies of dwarfs and giants. These corrections are discussed below and listed for all lines in online Table \ref{linelist}.

\paragraph{Oxygen.} \label{sec:nlteoxy} 
NLTE corrections for the O~I triplet were computed using the Kiel code
\citep{steen84} and MARCS model atmospheres. The O model atom is from
\citet{paunzen}. The cross-sections for inelastic H I collisions were computed
according to \citet{drawin}. Setting the scaling factor for these cross-sections
(S$_{\rm H}$) to $0$ and $1$, respectively, we obtain a correction of $-0.1$ dex in both cases. 

When we compare these corrections to the more recent values estimated from \citet{fabbian} we find that our corrections are lower. \citet{fabbian} estimated NLTE corrections for RR Lyr stars 500~K warmer than our stars and obtained values of -0.5 and -0.35 for S$_H$ = 0 and 1, respectively. Their Fig. 8 shows how the NLTE corrections for oxygen vary with temperature and gravity and demonstrates that temperature has a much larger influence on the corrections than gravity; a difference in $\Delta_{NLTE}$ for S$_H = 0$ 
of -0.4 dex corresponds to a difference of +1000~K in temperature. The warmer metal-poor stars yield a correction of -0.5 dex, compared to -0.1 dex for the cooler metal-poor turn-off stars. This would yield a correction of approximately -0.3 dex for our stars with S$_H$ = 0.

\paragraph{$\alpha$-elements.} 
\label{sec:nltealpha}  
For magnesium we use precise NLTE corrections kindly calculated by S. Andrievsky, using the MULTI and SYNTHV codes described in \citet{andrMg10}, taking fine structure of the 3p 3P* levels into account, and adopting our oscillator strengths. The accurate results for individual lines in our two stars are listed in the online Table \ref{linelist}. \dn \rm varies from 0.0 to 0.17 dex for the various Mg lines; an average value is $\sim$0.05 dex.

For Silicon, \dn is small, typically from 0.0 to +0.25 dex, according to \citet{Shi09}, who mainly consider dwarf stars; for the 390.5 nm line we adopt a mean value of 0.05 dex from the 14 stars of their Table 2.
\citet{preston} noted that silicon abundances for stars hotter than 5800~K decrease with increasing temperature and should thus not be trusted as chemical tracers. The temperatures of our stars are close to 5800~K, so we still trust our values (in LTE and NLTE), but note that a somewhat larger uncertainty ($\pm$ 0.1 dex) is probably in order.

Sulphur abundances from the 921.2 nm line need a larger correction. With 
(S$_{\rm H} = 1$), \dn = -0.4 dex is interpolated between the two sets of stellar parameters (\teff/\glog/[Fe/H] = 5500/2.0/-2.0 and 6500/2.0/-3.0) in \citet{SNLTEtakeda}.

For four of our nine Ca lines, \dn \rm has been taken from \citet{Mashon07}, who use a scaling factor of {\bf S$_{H}=0.1$}. The individual corrections are listed in Table \ref{linelist} and an average value in Table \ref{abuntab}.

No NLTE corrections currently exist for Ti.

\paragraph{Odd-$Z$ elements: Na, Al, Sc.} \label{sec:oddzlight}
We estimate \dn = -0.2 and -0.1 dex for the Na D1 and D2 lines in our RR Lyr stars, based on the calculations of \citep{AndrNa07}, who applied the same codes and methods as described for Mg above and Al below.
For Al, we estimate \dn = 0.7 dex for Al from \citet{AndrAl08}, using a scale factor of 0.1 in the \citet{drawin} formula. 

No \dn \rm values for Sc in metal-poor stars are available yet. \citet{ZhangSc} suggest a correction of 0.3 dex for the Sun (an average for the three lines seen in our RR Lyr stars). Since we are dealing with Sc II, one might expect to get small NLTE corrections. However, due to the very different stellar parameters of our RR Lyr stars and the Sun and the strong dependence of the Sc abundance on microturbulence, we choose to not apply their \dn \rm value to our stars.

\paragraph{Si-burning elements: Cr, Mn and Co.} \label{sec:siburn}
NLTE abundance corrections for Mn, and Co were taken from \citet{Berg10},
\citet{Berg08}, and \citet{BergCo}, respectively. The Cr correction is estimated to be around 0.4 dex (M. Bergemann, priv. comm.).

\paragraph{Neutron-capture elements: Sr, Ba.}\label{sec:rareearth} 
NLTE corrections for Sr in EMP stars are available from \citet{mash01}, who find \dn = 0.6 dex for stars of similar atmospheric parameters 
as ours, but do not list the lines used in this value. The correction is quite 
sensitive to the stellar parameters \citep{belSr, mash01}, but the above value 
should be a good estimate for our stars.

\dn for Ba is given by \citet{AndrBa09} for the three lines at 455.4 nm (around 0.15 dex, 585.3 nm, and 649.6 nm. Thus, only the resonance line (455.4 nm) can be corrected in our spectrum of CS 30317-056 and none in CS22881-039, so we prefer to retain the uncorrected LTE Ba abundances in both stars.

\begin{figure*}
\begin{center}
\includegraphics[width=0.4\paperwidth]{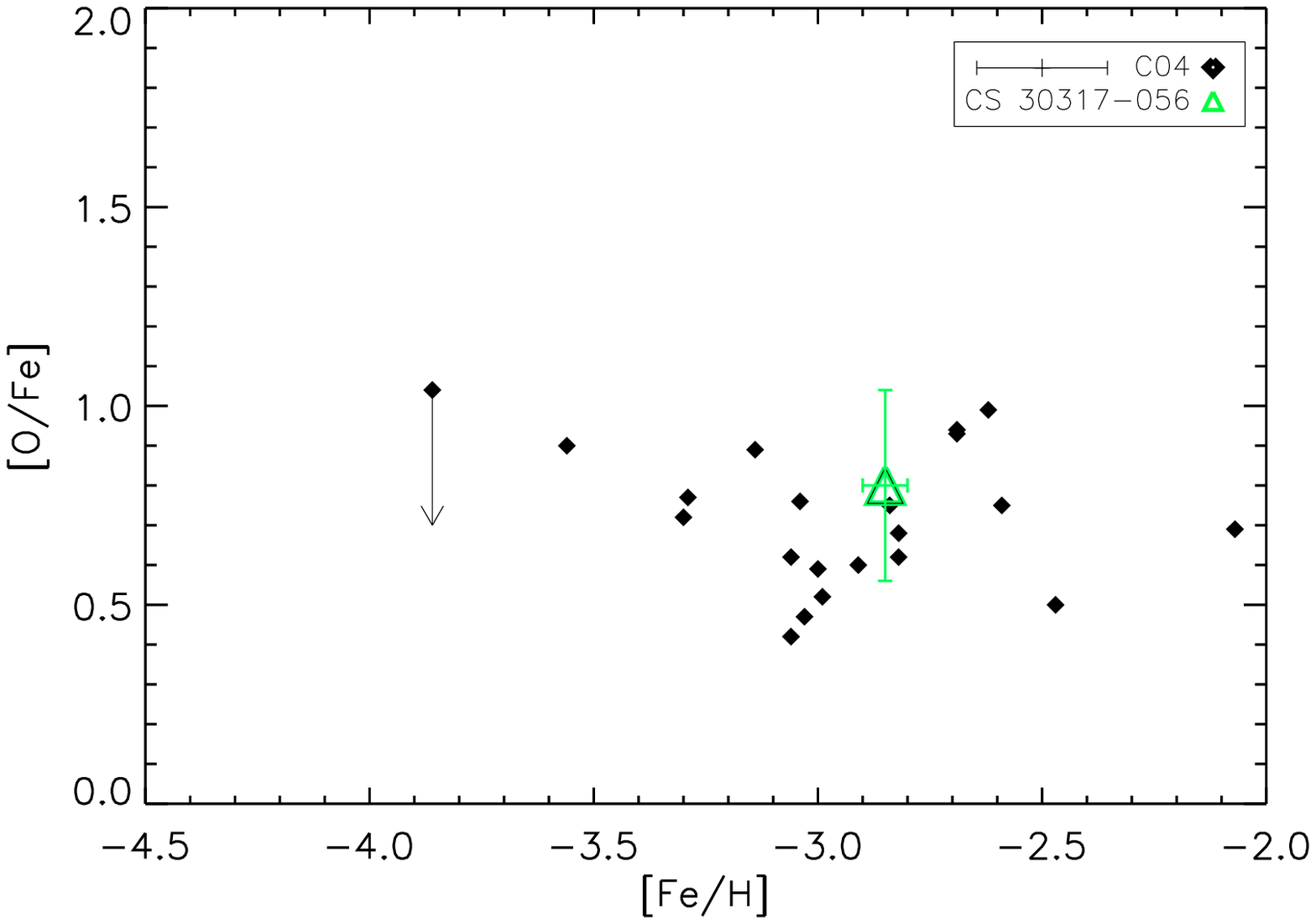}
\includegraphics[width=0.4\paperwidth]{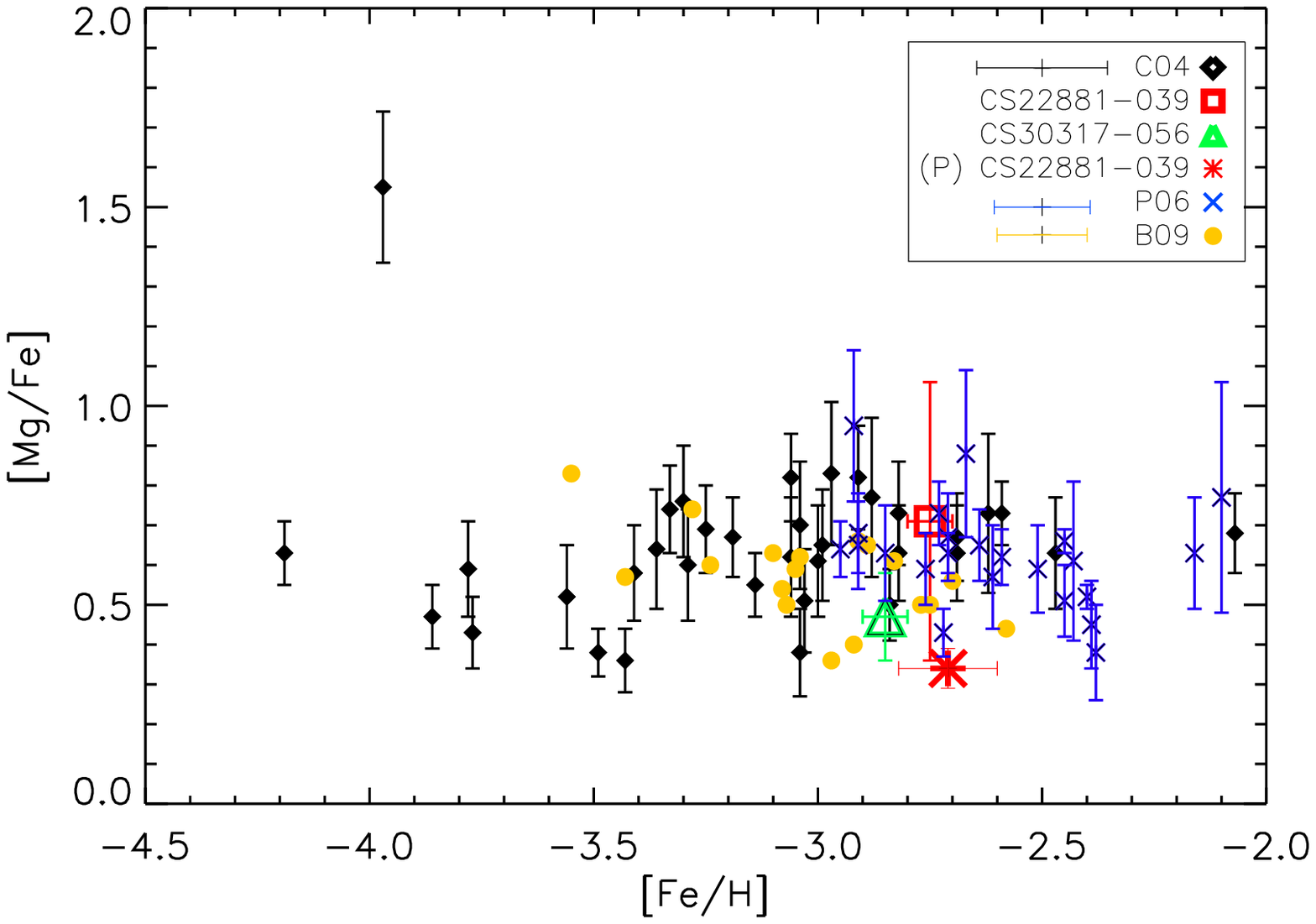}
\caption{Left: NLTE values of [O/Fe] for our RR Lyr stars compared to the giants from \citet{firstV}. Right: [Mg/Fe], corrected for NLTE, compared to EMP giants
\citep{firstV}, dwarfs \citep{Bonifaciodwarf}, as well as the RHB stars from \citep{preston}, all have been corrected for NLTE effects by applying corrections from \citet{andrMg10}. 
Results for CS~22881-039 are shown as a red square (this paper) and asterisk 
\citep{preston}; other symbols as in Fig. 2.}
\label{alpha}
\end{center}
\end{figure*}

\section{Results}

In this section we compare the abundances for our two RR Lyr stars with 
the earlier results for EMP giants and dwarfs \citet{firstV, firstVIII, Bonifaciodwarf}, 
and RHB stars \citep{preston}. Systematic differences in abundance scales between our results and the former should be minimal, because the observations and reductions were carried out in exactly the same way. Larger differences 
from the results of \citet{preston} might be expected, 
since different observational approaches, model atmospheres, codes, and methods were
used to determine the stellar parameters and abundances. Still, \citet{preston} find that differences in individual elemental abundances computed with
MARCS or ATLAS model atmospheres should be less than $0.05$ dex.

The abundance ratios relative to Fe are listed in Table \ref{abuntab} and shown graphically in the following (see e.g. Fig. \ref{alpha}, \ref{oddZ}, \ref{complete} and \ref{SrBafig}). All abundance ratios shown were obtained
assuming LTE, except when explicitly noted as corrected for NLTE effects. 
Note that NLTE effects in Fe might offset all these abundances ratios by a small amount.  This offset might be as large as 0.2 dex \citep{Mashonkina10} but these corrections still need inclusion of many high energy levels to predict accurate abundance corrections. 
Generally, the figures show an abundance y-range of 2 dex, except for O, Sr and Ba. The two exceptional stars CS~22949-037 \citep{firstV} and CS~29527-15 
\citep{Bonifaciodwarf} have been omitted in the comparisons. Overall, the 
detailed chemical composition of our two RR Lyr stars does not deviate significantly 
from the well-known mean trends for ``normal'' EMP giant and dwarf stars; 
we comment on individual element groups below.

The element abundances derived from the three spectra listed in Table \ref{abfe} are very similar and yield very homogeneous abundances (the differences in abundances from the individual spectra are $\sim$ 0.05 dex - per element). Hence, the abundances are stable during the different pulsation phases (as noted in \citet{kolenberg}), and should therefore be considered trustworthy as chemical tracers.
Furthermore, the coadded spectra yield abundances agreeing within 0.05 dex with those obtained from the single spectra, except for a few elements, where the difference is around 0.1 dex.

\subsection{Lithium}
As expected, Li is not observed in our RR Lyr stars. During the evolution of the star along the RGB, the increasingly deep mixing has filled the external layers with the matter of deep, hot layers where lithium has been destroyed by proton fusion.

\subsection{The $\alpha$-Elements}

As in other EMP stars, we find [$\alpha$/Fe] $>0$ in our targets, with relatively small dispersion. We discuss the individual elements below.

\noindent\textit{Oxygen}\rm \\
O was only detected in CS~30317-056. Its abundance was determined from the 
O~I triplet at 777.194 nm, which is affected by NLTE, while the O abundance
for giant stars in \citet{firstV} were based on the forbidden [O I] line at 630
nm, which is insensitive to NLTE effects, but may be affected by 3D effects. However the O I 3D effects remain small for giants \citep{collet}. 
As Fig. \ref{alpha} shows, especially the 
corrected value of [O/Fe] in CS~30317-056 (Sect. \ref{nlte}) agrees well with the mean value for EMP giants found by \citet{firstV} without regard to 3D effects.\\

\begin{figure}[!htp]
\begin{center}
\includegraphics[width=0.4\paperwidth]{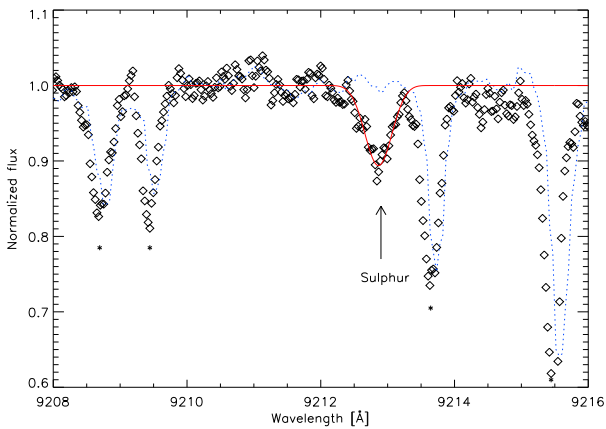}
\caption{The observed spectrum of CS~30317-056 around the S line 921.3 nm 
(open diamonds). The dotted blue line shows the spectrum of a fast-rotating star with mainly telluric lines (indicated by '*'). This confirms that the S line is indeed stellar and not telluric. The solid red line is an LTE synthetic spectrum fit to the S line, which confirms the abundance we obtain from direct integration.}
\label{Sulphur}
\end{center}
\end{figure}

\noindent\textit{Mg, Si, S, Ca, and Ti}\rm \\
Our Mg abundances are based on six lines in both stars, In CS~22881-39 the strong magnesium lines are asymmetric, which could indicate strong velocity fields inside the atmosphere (the star has been observed just before the sudden rise in luminosity). As a consequence, the Mg abundance deduced from a static model atmosphere cannot be completely reliable.

Small positive NLTE corrections have been 
applied to all Mg abundances in Fig. \ref{alpha}. For our RR Lyr stars (see Table \ref{abuntab}) they were calculated specifically for this study (Sect. \ref{sec:nltealpha}); for the \citet{preston} HB stars we estimate \dn = 0.2 -- $0.3$ dex, depending on the 
stellar parameters. This correction may be overestimated, since we found a much smaller value when getting the exact calculations compared to estimating the correction based on \citet{andrMg10}. 

Our LTE abundances of Si are based on the single line at 390.5 nm and exhibit large
scatter, as also found by \citet{preston}. Our abundances are somewhat lower 
than those for giants by \citet{firstV}, but agree with those for turnoff stars 
by \citet{Bonifaciodwarf}. This difference may be linked to the temperature dependence noted by \citet{preston}, who recommended to only trust abundances for stars with temperatures below 5800~K. Our stars are close to this limit, so we expect our Si abundances to be fairly reliable. (The difference between the dwarfs and giants could be due to NLTE effects).  

Sulphur is an interesting element, because it has so far only been detected in a few EMP stars (e.g., CS~29497-030, \citet{firstIV}). Here it is measurable in 
CS~30317-056 only (see Fig. \ref{Sulphur}). An LTE analysis of the strong 
triplet at 921.286 nm yields [S/Fe] = 0.78 dex, above the mean 
value by \citet{svovlnis}, but in agreement with \citet{cafsvovl} and 
other sources listed in \citet{svovlnis}. Using the value of \dn = -0.4 dex 
from \citet{SNLTEtakeda}, the agreement is essentially perfect.

Our [Ca/Fe] ratios from six Ca I lines, whether derived in LTE or NLTE, agree 
within the errors with \citet{preston} and \citet{firstV}. The LTE abundance of Ti in 
CS~22881-039 match the Ti abundances for giants \citep{firstV}, 
while that for CS~30317-056 matches the higher values for dwarfs by \citet{Bonifaciodwarf}; both are in the range found for EMP RHB stars by \citet{preston}. 
The Ti~I and Ti~II LTE abundances agree well, within 0.08 dex, supporting the gravity derived from the ionisation balance of Fe.

\begin{figure}[!htp]
\begin{center}
\includegraphics[width=0.4\paperwidth]{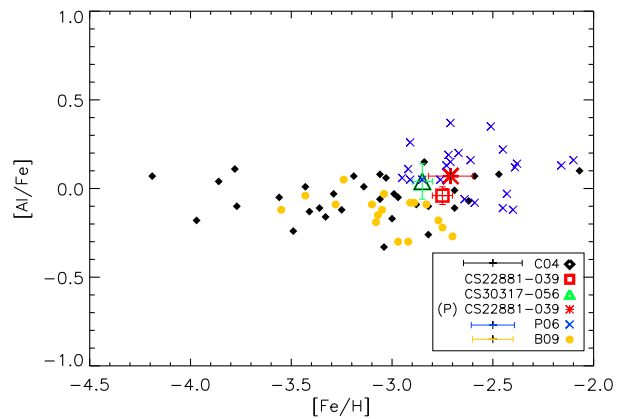}
\caption{[Al/Fe] for our RR Lyr stars and the RHB stars by \citet{preston}, as well as for giants \citet{firstV}, all corrected for NLTE; symbols as in Fig. \ref{alpha}.}
\label{oddZ}
\end{center}
\end{figure}

\subsection{The Odd-Z Elements Na, Al, and Sc}

Na and Al abundances derived from the resonance lines of the neutral atoms are very sensitive to NLTE effects. 
According to \citet{firstV} and \citet{Bonifaciodwarf}, \dn is of the order of
$-0.2/ -0.1$ dex for Na and $+0.70$ dex for Al. Fig. \ref{oddZ} compares our NLTE results for Al with those from the earlier study by \citep{AndrAl08}. From the stellar parameters and the calculations of \citet{AndrAl08}, we estimate \dn for the \citet{preston} RHB stars to be 0.8 -- 1.0 dex for Al, depending on the stellar parameters. 

The values of [Sc/Fe] for both RR Lyr stars are well within the range seen in \citet{firstV}. The observed trend of [Sc/Fe] is flat, with
a scatter of only $0.12$ dex \citep{firstV}.

\begin{figure}
\begin{center}
\includegraphics[width=0.4\paperwidth]{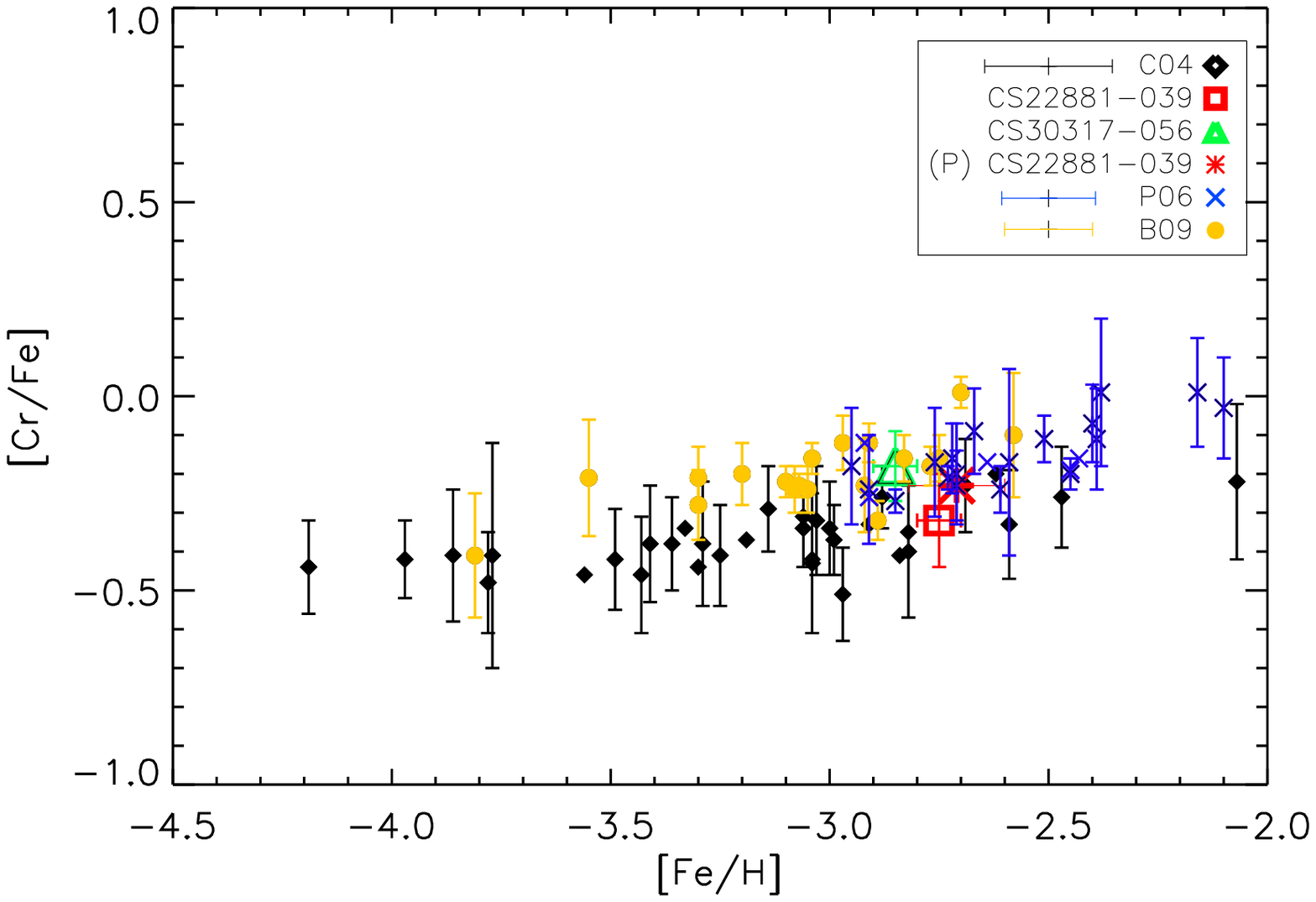}
\includegraphics[width=0.4\paperwidth]{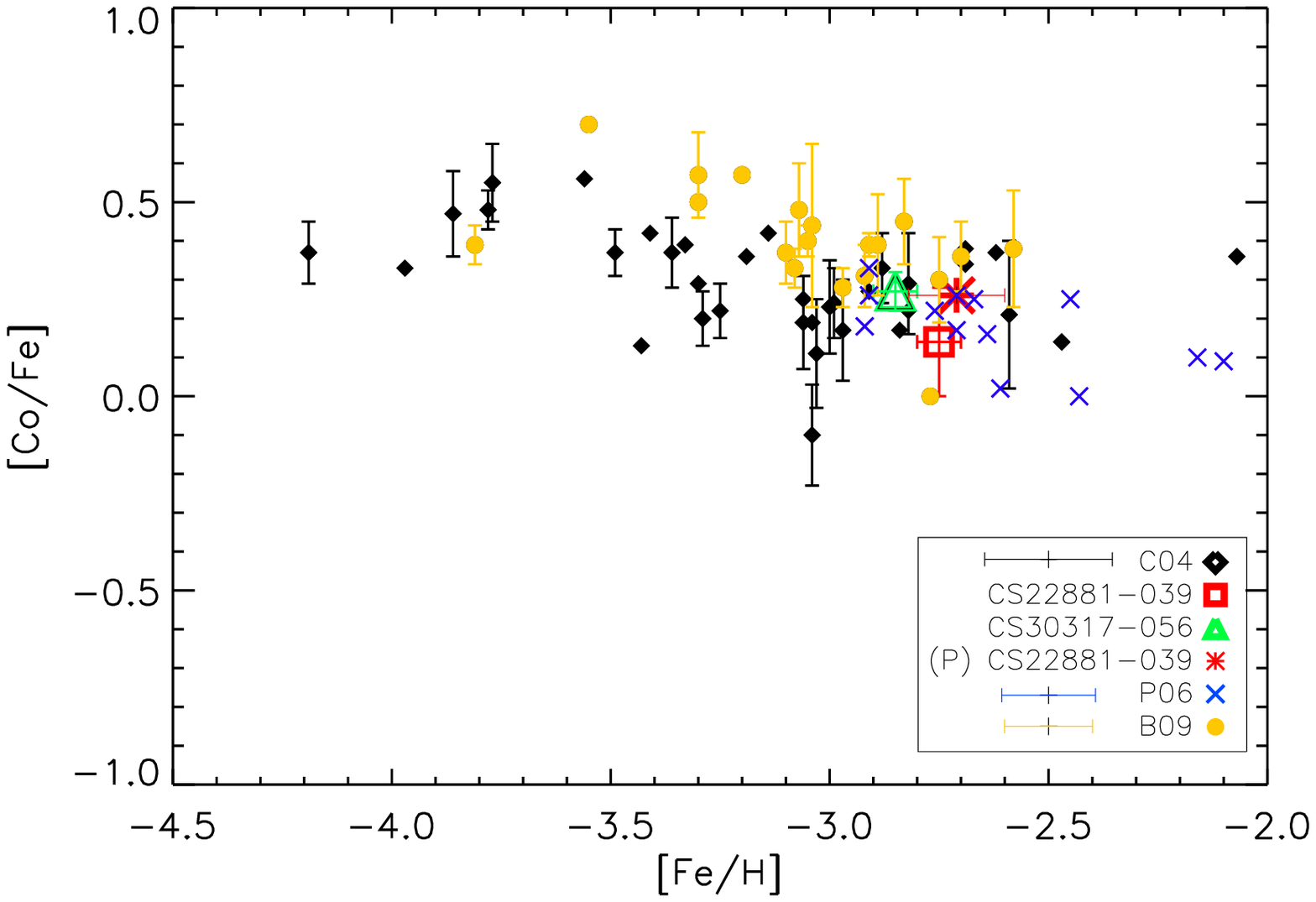}
\caption{[Cr/Fe] and [Co/Fe] vs. [Fe/H] (symbols as in Fig. \ref{alpha}), all derived in LTE. The RR Lyr stars agree well with the results of \citet{preston}, \citet{firstV} and \citet{Bonifaciodwarf}. }
\label{incomplete}
\label{complete}
\end{center}
\end{figure}

\subsection{Iron-peak Elements}

To facilitate comparison with theoretical yields from supernova models,
we discuss the two subgroups of the iron-peak elements separately. We note 
that iron itself, which is used as the reference element, may be affected by NLTE effects \citep{Mashonkina10}, which would shift all the abundance ratios discussed in the following slightly up or down by a fixed amount.
Most of the iron-peak abundance ratios are extremely tightly defined -- so
tightly, in fact, that interpretations of the slopes (notably that of the
[Cr/Fe] relation) in terms of metallicity-dependent supernova masses appear
implausible. \citet{firstV} discussed residual NLTE effects for the abundance
trends as a possible alternative, which seems to be confirmed by the diverging
abundances from Cr~I and Cr~II lines found by \citet{Bonifaciodwarf}.\\

\noindent\textit{Incomplete Silicon Burning Elements: Cr and Mn}\\
Our Cr abundances are based on three Cr~I lines, and the Mn abundance on the three Mn~I resonance lines near $403$ nm (see Table \ref{linelist}). Since the Mn lines are weak in these stars, no special treatment of the hyperfine
structure was necessary. No lines of ionised Cr and Mn were detected in
our RR Lyr stars.

The LTE abundances of Cr and Mn for our two RR Lyr stars agree within 0.1 dex with those by \citet{preston}, \citet{firstV}, and \citet{Bonifaciodwarf} (Fig. \ref{incomplete}) -- better than the combined errors. \citet{Bonifaciodwarf} suggest that NLTE effects in Cr may be significant in metal-poor stars; if so, [Cr/Fe] in this metallicity range should be $\sim0.1$.\\

\noindent\textit{Complete Silicon Burning Elements: Co and Ni}\\
Our Co and Ni abundances are based on two neutral lines each in both stars. 
Within our estimated error, we find good agreement between our LTE abundances and those of the EMP stars of \citet{firstV}, \citet{Bonifaciodwarf}, and \citet{preston} (Co only). However, we note that NLTE effects for Co in our RR Lyr stars are large and positive (Table \ref{abuntab}). \\

\subsection{Neutron-Capture Elements: Sr and Ba}

\begin{figure}[!htp]
\vspace{-7mm}
\begin{center}
\includegraphics[width=0.42\paperwidth]{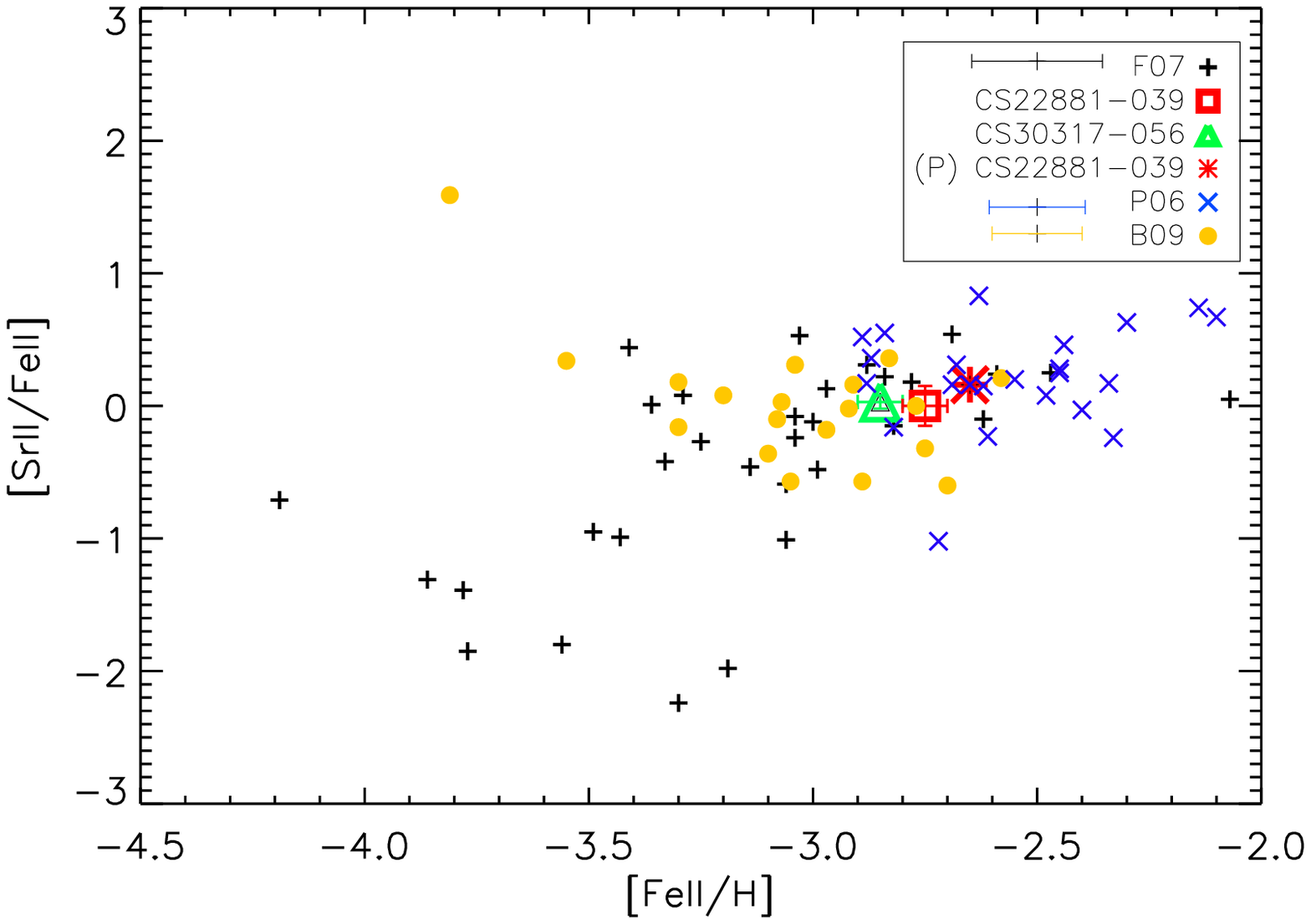}
\includegraphics[width=0.42\paperwidth]{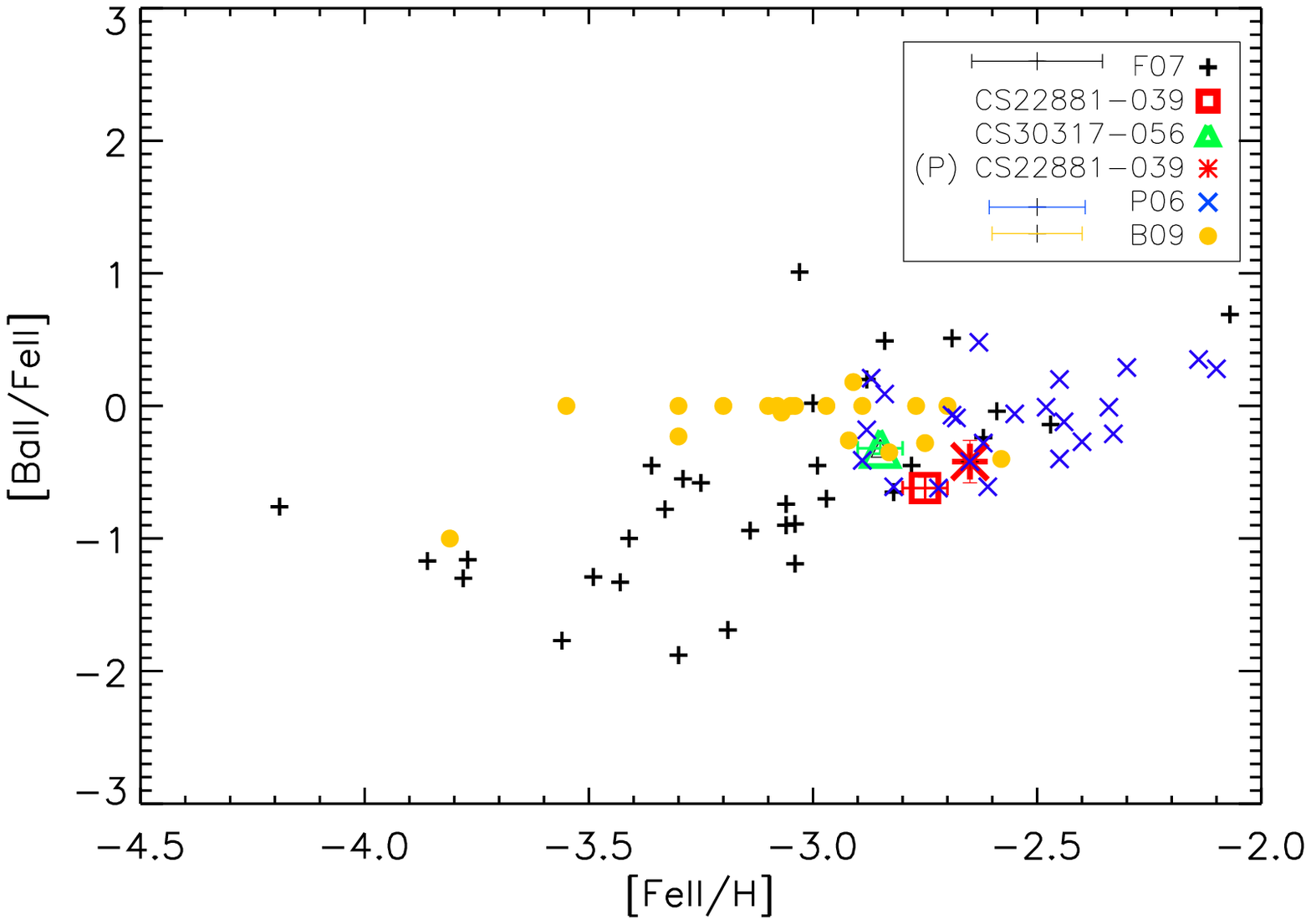}
\caption{[Sr/Fe] and [Ba/Fe] in EMP stars, assuming LTE (symbols as in 
Fig. \ref{alpha}; note the change in vertical scale here). 
Our RR Lyr stars agree well with the previous relations for giants, dwarfs and 
RHB stars \citep{firstV,Bonifaciodwarf, preston}. }
\label{SrBafig}
\end{center}
\vspace{-5mm}
\end{figure} 

Sr and Ba are the only neutron-capture elements detected in our RR Lyr stars. 
Their abundances are determined from the resonance lines and hence
subject to NLTE effects. However, as noted above, consistent NLTE corrections for Ba in our stars are not available in the literature, so we have chosen not to correct our [Ba/Fe] ratios for NLTE.

Both [Sr/Fe] and [Ba/Fe] show huge scatter below [Fe/H]$ \sim-3$, far in 
excess of observational errors \citep{prim94, McWilliam, firstVIII} -- see 
Fig. \ref{SrBafig}. Taking this into account, our results for CS~22881-039 
and CS~30317-056 (both corrected and uncorrected) agree within the combined errors with 
the abundances found by \citet{preston} and \citet{FS10} for the RHB stars and also with the concordant results for EMP dwarfs and giants discussed by
\citet{Bonifaciodwarf}. 

\section{Supernova Models and Yields}\label{SNmass}

Having verified that our RR Lyr stars can be included in general samples of EMP
stars, we proceed to compare the derived stellar abundances in our EMP RR Lyr stars with the supernova model yields of \citet{limongi},
\citet{galchem}, \citet{SN13-50} and \citet{Izu}. It is generally believed
that SNe of Type Ia did not contribute significantly to the composition of the EMP stars, so only models of core-collapse SNe with massive progenitor stars need to be considered here.

These models are characterised by a number of parameters, such as the mass of the progenitor, the explosion energy, the peak temperature, the mass cut, the electron density in the proto-neutron star, $Y_e$ {\bf \footnote{$Y_e$, is the so-called electron fraction, which describes the number of electrons per nucleon.}}, the amount of fallback and/or mixing, and the degree of anisotropy, including any pre-explosion jets. Variations in these parameters are manifested in characteristic changes in the predicted element ratios, so the observed abundance patterns can be used to constrain the values of these parameters. For such comparisons, NLTE corrected abundances must of course be used whenever available -- see Table \ref{abuntab}.

Even EMP stars are presumably the result of more than one SN explosion, so 
the assumed properties of the progenitor population -- notably the IMF -- enter implicitly into the comparison of models vs. observations. Only models assuming a Salpeter IMF have been chosen here in order to compare models
on an equal basis without introducing differences due to changes in the IMF. 

A Salpeter IMF is the standard assumption in supernova and galactic chemical evolution models. Top-heavy IMF's have been tested as well by \citet{ballero, matteucci}. Both studies find that if the presumed POP III stars more massive than 100 M$_{\odot}$ resulting in Pair Instability SNe (PISNe) with yields showing a strong odd/even effect, the overall impact on the early ISM is negligible. Only if these very massive stars kept forming for generations would one see an imprint of them, and this is not detectable in the observations. In fact, \citet{ballero} find that a constant or slightly varying ISM yields the best results when comparing to observations of metal-poor stars.

\begin{figure}
\begin{center}
\includegraphics[width=0.5\textwidth,angle=0.0]{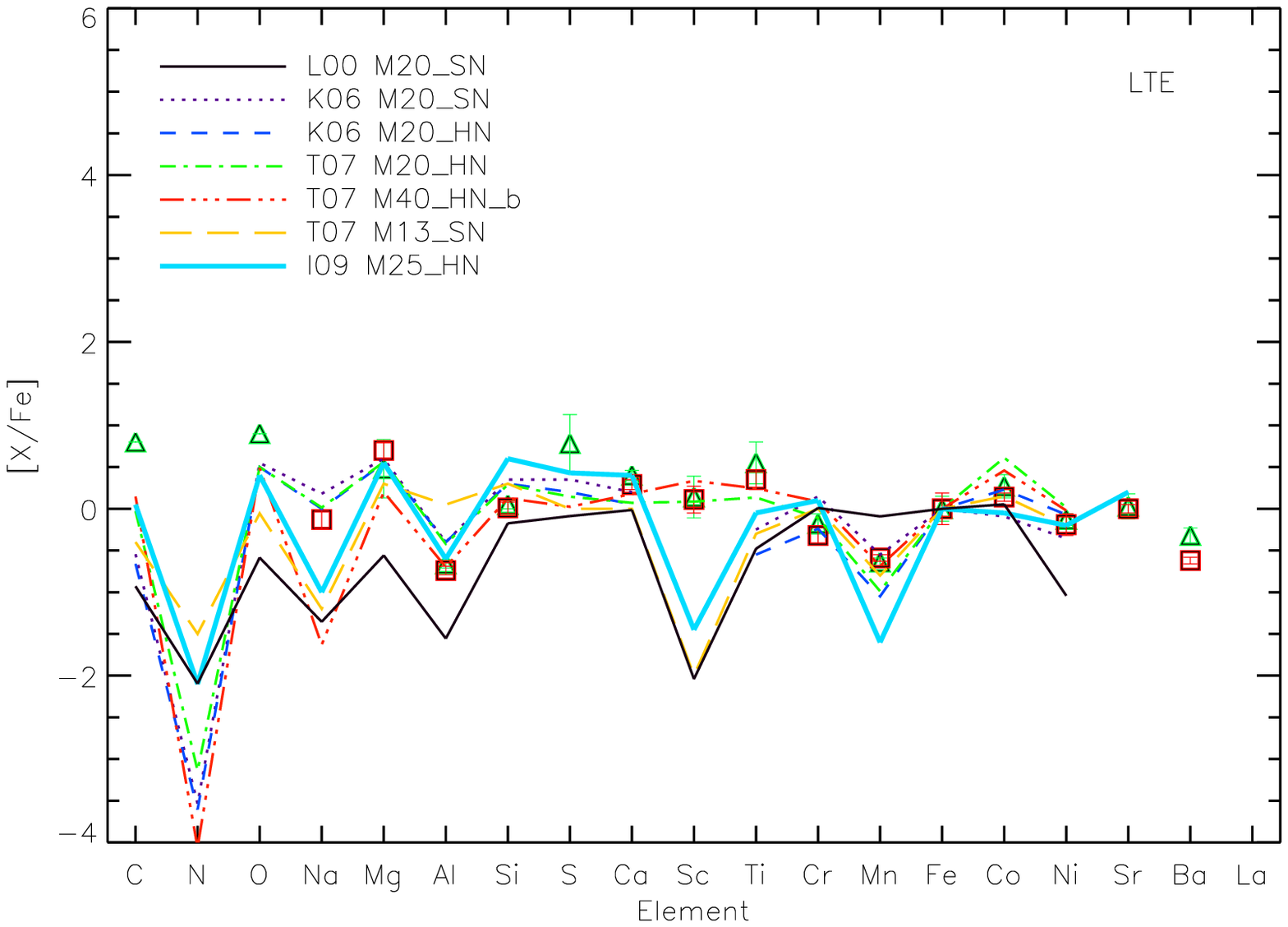}
\includegraphics[width=0.5\textwidth,angle=0.0]{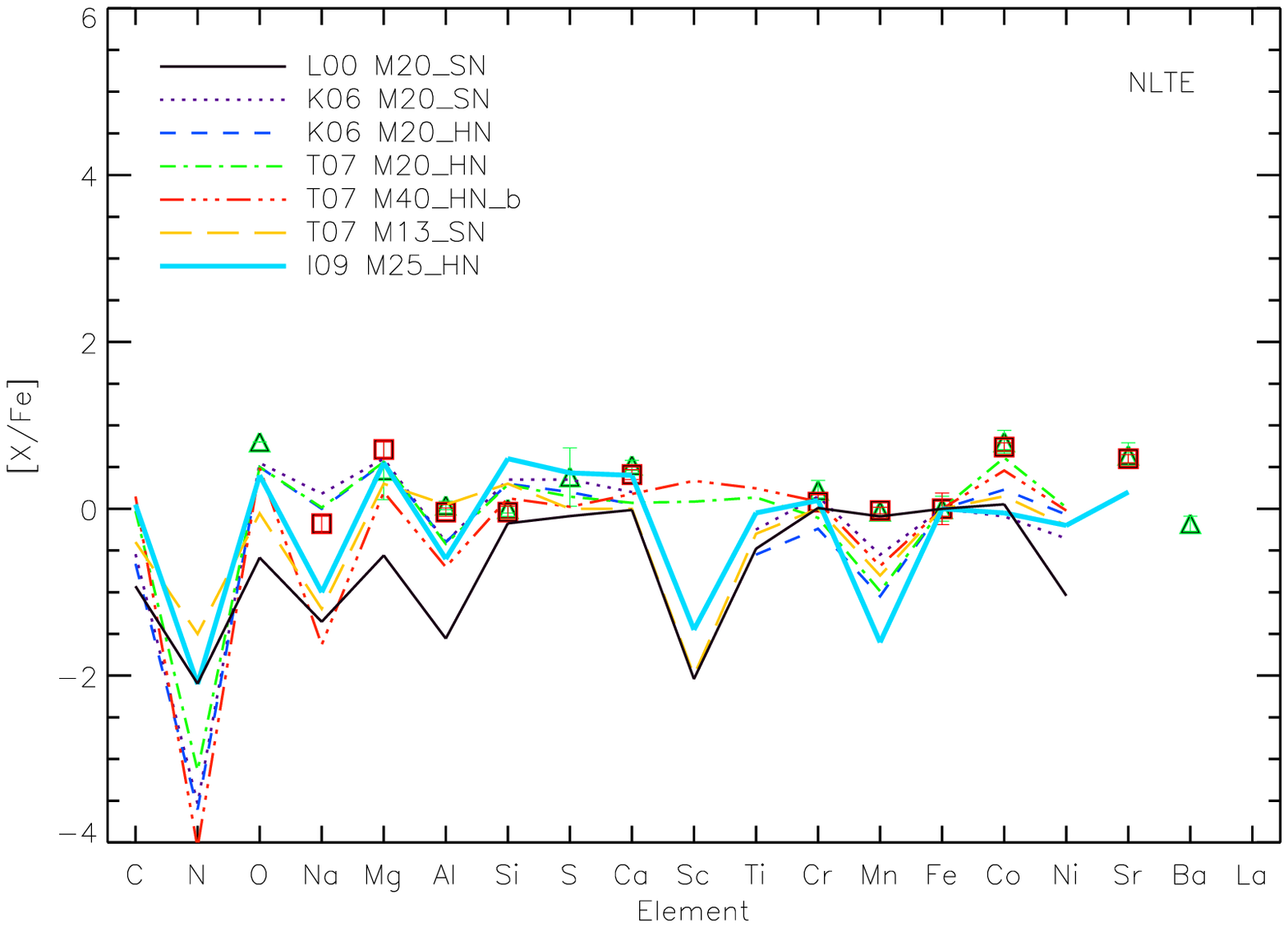}
\caption{Observed abundance ratios in our RR Lyr stars from Table \ref{abuntab}
(symbols as in Fig. \ref{alpha})vs. predicted yields for models of super- and
hypernovae (HN) by: \citet{limongi} (full black line), \citet{galchem} (dotted:
SN; dashed: HN), \citet{SN13-50} (dot-dashed and triple-dot-dashed: 20 and 40
$M_{\odot}$ HN; long-dashed: 13 $M_{\odot}$ SN) and \citet{Izu} 25 $M_{\odot}$ HN; thick blue solid line. }
\label{SNmodelyield}
\end{center}
\end{figure}

Fig. \ref{SNmodelyield} shows that the various model yields differ most strongly for certain elements, which therefore become the most important diagnostics of the different model features. These elements are, notably, Na, Mg, Al, Sc, Mn and the neutron-capture element Sr. \citet{lai08} found that Si, Ti, and
Cr exhibit clear trends with stellar parameters such as the temperature, so these elements are less useful in this context. 

Only models that provide the fit to the abundances observed are included in Fig. \ref{SNmodelyield}. 
Models of the putative, extremely massive PISNe \citep{heger02,firstII}) predict a much larger odd-even effect than observed. These very heavy SN were therefore not considered here.

\subsection{The $\alpha$-elements}

The $\alpha$-elements are generally formed in massive progenitor stars already before the SN explosion and thus bring information on the nature of the first stellar population(s). E.g., O (and C) are primarily produced in core helium burning in massive stars, O also during Ne burning, and Mg is produced in shell carbon burning. The paucity of iron formed by SNe II relative to the Solar composition causes the familiar over-abundance of the $\alpha$-elements, typically $\sim$0.3 dex, in metal-poor stars (see Table \ref{abuntab}). 

Supernova models \citet{heger+woosley, galchem} indicate that larger yields of even-Z elements, also seen as a large odd-even effect, correspond to a larger (total or at least envelope) mass of the progenitor star. However, the magnitude of the observed odd-even effect depends heavily on a few relatively large NLTE abundance corrections, especially that for Al, so computing or estimating \dn correctly is of direct importance for the conclusions that can be drawn on the mass of the progenitor stars. 

\subsection{The Odd-Z Elements}

The odd-Z elements are generally under-abundant by $\sim$0.2 dex relative to the Sun. Na and Al are produced in the progenitor before the SN explosion, but Sc is produced in explosive O and Si burning, and also in Ne burning \citep{WW95}, hence in the SN itself. 

The Sc abundance appears to be the best diagnostic of the explosive energy of the SN: The high (=near-Solar) [Sc/Fe], and to a lesser degree [Ti/Fe], ratios suggest that very energetic SNe (Hypernovae) were the progenitors of the EMP stars we observe today. We note, however, that \dn for Sc has not yet been computed, and a large negative value would affect this conclusion.

Moreover, according to \citet{Izu}, high Sc and Ti are tracers of a high $Y_e$ to an even larger extent than a high explosion energy.  A high-energy explosion is still required, but it is not the primary cause of a large production of 
Sc and the neutron capture elements. Models with (artificially) high values of $Y_e$ \citep{Izu, SN13-50} and low-density modifications caused by, e.g., jet-like explosions and fallback \citep{SN13-50} fit the abundance patterns of EMP stars better, as indicated by the relatively large amounts of Sc and Sr. The caveat about NLTE effects should be kept in mind, however.

\subsection{Silicon Burning Elements}

Among the iron-peak elements, Cr and Mn are produced by incomplete silicon
burning and Co, Ni and Zn in complete silicon burning during supernova explosions. Incomplete silicon burning occurs at temperatures
of $4\cdot10^9 $K $<T_{peak}<5\cdot10^9 $K \citep{nomotoSNHN}, complete
silicon burning at temperatures above $5\cdot10^9 $K.
The observed high abundances of the complete Si burning elements indicate that the peak temperatures reached by the progenitor SNe was indeed high. 
At higher metallicities than discussed here, the production of Ni (and Fe) is, of course, dominated by the entirely different processes in SNe Ia. 

\subsection{Neutron-capture Elements}\label{ncapelem}

At Solar and moderately low metallicity, the abundances of Sr and Ba are dominated by the s-process \citep{Arland}. In EMP stars, however, the 
neutron-capture elements should all be due to the r-process
\citep{oldspite,Tru,firstVIII}, so observations of Sr and Ba in EMP stars should provide useful constraints on the site of the $r$-process \citep{heger+woosley}.

In past decades, core-collapse supernovae have been considered the most promising astrophysical site for the r-process, but this scenario is facing difficulties \citep[e.g.][]{wan2010}, and there is presently no consensus on the site of the $r$-process(es). 

Unfortunately, we only have NLTE-corrected abundances for Sr, and \dn for Ba might be appreciable (see Sect. \ref{sec:rareearth}), so our present ability to 
place strong constraints is limited.
Constraints on $Y_e$ could be derived from abundances of light neutron-capture elements, i.e. elements in the range $38 < Z < 48$. Slight enhancements of Sc (and Ti) are consistent both with a high $Y_e$ or a jet-like explosion (cf. the good fit of the \citet{SN13-50} models to the data in Fig. \ref{SNmodelyield}). 

\section{\teff at the RR Lyr-RHB Boundary}
One of the aims of the paper of \citet{preston} was to determine the temperature of the red edge of the RR Lyr instability strip (the ``fundamental red edge'': FRE in the pulsation theory). 
Horizontal branch stars cooler than the FRE do not pulsate, because at this temperature the onset of convection disrupts the pulsation mechanism. This limit is difficult to predict theoretically, however, because such a prediction requires a very good understanding of the role of convection in RR Lyr pulsations. \citet{preston} estimated the FRE temperature by comparing the temperature distribution of RR Lyr stars and RHB stars in globular clusters and in field stars. He found that in both cases no RR Lyr star was found below log \teff = 3.80. Accordingly, he adopted for $\rm [Fe/H] < -2$,  
$\rm log T_{eff}(FRE) = 3.80 \pm 0.01$ or $\rm T_{eff}(FRE) = 6300 \pm 150~K$.
 
CS~22881-039 and CS~30317-056 are both significantly cooler than the FRE as defined by \citet{preston}; nevertheless, they are photometric variables with light curves and periods like those of RR Lyr stars. \citet{Nem04} found three RR Lyr below this limit in NGC5053 ($\rm[Fe/H]\sim -2.30$), and it is interesting that the photometric amplitude of these stars is small ($\rm <0.6 mag$).
In a recent paper, \citet{FS10} compared the temperature distributions of 
non-variable HB stars and RR Lyr stars and estimated the temperature of the FRE in the interval $\rm -0.8 \lesssim [Fe/H] \lesssim -2.5$ be about 5900~K, in good agreement with our result. Hence, at the lowest metallicities the temperature of the FRE is probably lower than estimated by \citet{preston}; ${\rm T_{eff}(FRE)} \approx $ 5900 K would be probably a better estimate. Moreover, the RR Lyr stars found very close to the FRE seem to have smaller photometric amplitudes than typical RRab.

\section{Conclusions}

Our detailed analysis of two EMP RR~Lyr stars from high-resolution VLT/UVES spectra
have yielded abundances for 16 elements, which agree very well (generally within $\leq 0.2$ dex) with those by \citet{firstV}, \citet{Bonifaciodwarf}, and \citet{preston}, for giants, dwarfs, and RHB stars, respectively. 
Due to the homogeneity of the stellar abundances from the individual spectra, despite the pulsation, our RR Lyr stars can be used as chemical tracers as well as non-variable RHB stars.

The temperature of the fundamental red edge of the instability strip is probably lower at low metallicity than obtained by \citet{preston}; our results suggest a value of $T_{\rm eff}(FRE) \approx 5900~K$, in agreement with \citet{FS10}.

Our comparison of the abundance patterns of well-studied EMP stars with supernova yields from the models of \citet{SN13-50,nomotoSNHN} with varying $Y_e$, mass cut, mixing, and fallback indicates that supernovae with
relatively large masses (up to 40 M$_{\odot}$) and large explosion
energies provide the best overall match to the observations. Including the 
NLTE corrections leaves much the same overall picture of the predecessor
SNe, but lowers the estimate of the progenitor mass.

Overall, among current models, the 40 M$_{\odot}$ HN model by \citet{SN13-50}
seems to provide the best fit to our LTE abundance trends, even though it
is not straightforward to constrain the mass and energy of the previous
generation of SNe when varying some parameters such as $Y_e$ while keeping others fixed. 
On the other hand, when we compare the NLTE corrected abundances to the same
models, the 25 $M_{\odot}$ hypernova model by \citet{Izu} appears to give the best fit (or the 20 $M_{\odot}$ HN model by \citet{SN13-50}) , because the NLTE abundances of such elements as Na, Al, S, Sc, Ti and the iron-peak elements differ from the predictions of the 40 $M_{\odot}$ HN model. Hence, good determination of the NLTE effects for these elements are particularly important in comparisons with SN models. Sc is a key element in such work, and a good NLTE analysis of Sc is an urgent priority. Zn is another very important diagnostic element, but is not observable in such hot EMP stars as those discussed here.

\begin{acknowledgements}
We acknowledge the help of Professor Andrew Cameron in determining the
photometric phases for CS 22881-39 and CS 30317-56, using light curves
from the Wide-Angle Search for Planets (WASP) project. The WASP consortium
comprises the Universities of Keele, Leicester, St Andrews, Queen's University Belfast, the Open University and the Isaac Newton Group. Funding for WASP comes from the consortium universities and from the UK's Science and Technology Facilities Council.
We acknowledge partial support from the EU contract MEXT-CT-2004-014265 (CIFIST). TCB acknowledges partial support from a series of grants from the US
National Science Foundation, most recently AST 04-06784, as well as from grant
PHY 02-16783 and PHY 08-22648: Physics Frontier Center/Joint Institute for
Nuclear Astrophysics (JINA). BN and JA thank the Carlsberg Foundation and the
Danish Natural Science Research Council for financial support. CJH thanks ESO for partial support, and N. Tominaga for providing some of the SN models used in this paper. S. Andrievsky kindly computed \dn values for Mg for this paper.
This research has made use of NASA's Astrophysics Data System. This research has
made use of the SIMBAD database, operated at CDS, Strasbourg, France, and the Two Micron All Sky Survey, which is a joint project of the University of Massachusetts and the Infrared Processing and Analysis Center/California Institute of Technology, funded by the National Aeronautics and Space Administration and the National Science Foundation. 

\end{acknowledgements}


\begin{thebibliography}{67}
\expandafter\ifx\csname natexlab\endcsname\relax\def\natexlab#1{#1}\fi

\bibitem[{{Alvarez} \& {Plez}(1998)}]{turbospectrum}
{Alvarez}, R. \& {Plez}, B. 1998, \aap, 330, 1100

\bibitem[{{Andrievsky} {et~al.}(2010){Andrievsky}, {Spite}, {Korotin}, {Spite},
  {Bonifacio}, {Cayrel}, {Fran{\c c}ois}, \& {Hill}}]{andrMg10}
{Andrievsky}, S.~M., {Spite}, M., {Korotin}, S.~A., {et~al.} 2010, ArXiv
  e-prints

\bibitem[{{Andrievsky} {et~al.}(2007){Andrievsky}, {Spite}, {Korotin}, {Spite},
  {Bonifacio}, {Cayrel}, {Hill}, \& {Fran{\c c}ois}}]{AndrNa07}
{Andrievsky}, S.~M., {Spite}, M., {Korotin}, S.~A., {et~al.} 2007, \aap, 464,
  1081

\bibitem[{{Andrievsky} {et~al.}(2008){Andrievsky}, {Spite}, {Korotin}, {Spite},
  {Bonifacio}, {Cayrel}, {Hill}, \& {Fran{\c c}ois}}]{AndrAl08}
{Andrievsky}, S.~M., {Spite}, M., {Korotin}, S.~A., {et~al.} 2008, \aap, 481,
  481

\bibitem[{{Andrievsky} {et~al.}(2009){Andrievsky}, {Spite}, {Korotin}, {Spite},
  {Fran{\c c}ois}, {Bonifacio}, {Cayrel}, \& {Hill}}]{AndrBa09}
{Andrievsky}, S.~M., {Spite}, M., {Korotin}, S.~A., {et~al.} 2009, \aap, 494,
  1083

\bibitem[{{Arlandini} {et~al.}(1999){Arlandini}, {K{\"a}ppeler}, {Wisshak},
  {Gallino}, {Lugaro}, {Busso}, \& {Straniero}}]{Arland}
{Arlandini}, C., {K{\"a}ppeler}, F., {Wisshak}, K., {et~al.} 1999, \apj, 525,
  886

\bibitem[{{Asplund} {et~al.}(2009){Asplund}, {Grevesse}, {Sauval}, \&
  {Scott}}]{asp09}
{Asplund}, M., {Grevesse}, N., {Sauval}, A.~J., \& {Scott}, P. 2009, \araa, 47,
  481

\bibitem[{{Asplund} {et~al.}(1997){Asplund}, {Gustafsson}, {Kiselman}, \&
  {Eriksson}}]{OSMARCS}
{Asplund}, M., {Gustafsson}, B., {Kiselman}, D., \& {Eriksson}, K. 1997, \aap,
  318, 521

\bibitem[{{Ballero} {et~al.}(2006){Ballero}, {Matteucci}, \&
  {Chiappini}}]{ballero}
{Ballero}, S.~K., {Matteucci}, F., \& {Chiappini}, C. 2006, \na, 11, 306

\bibitem[{{Beers} {et~al.}(1985){Beers}, {Preston}, \& {Shectman}}]{Beers85}
{Beers}, T.~C., {Preston}, G.~W., \& {Shectman}, S.~A. 1985, \aj, 90, 2089

\bibitem[{{Beers} {et~al.}(1992){Beers}, {Preston}, \& {Shectman}}]{Beers1992}
{Beers}, T.~C., {Preston}, G.~W., \& {Shectman}, S.~A. 1992, \aj, 103, 1987

\bibitem[{{Belyakova} \& {Mashonkina}(1997)}]{belSr}
{Belyakova}, E.~V. \& {Mashonkina}, L. 1997, Astronomy Reports, 41, 291

\bibitem[{{Bergemann} \& {Gehren}(2008)}]{Berg08}
{Bergemann}, M. \& {Gehren}, T. 2008, \aap, 492, 823

\bibitem[{{Bergemann} {et~al.}(2009){Bergemann}, {Pickering}, \&
  {Gehren}}]{BergCo}
{Bergemann}, M., {Pickering}, J.~C., \& {Gehren}, T. 2009, \mnras, 1703

\bibitem[{{Bergemann} {et~al.}(2010){Bergemann}, {Pickering}, \&
  {Gehren}}]{Berg10}
{Bergemann}, M., {Pickering}, J.~C., \& {Gehren}, T. 2010, \mnras, 401, 1334

\bibitem[{{Bonifacio} {et~al.}(2009){Bonifacio}, {Spite}, {Cayrel}, {Hill},
  {Spite}, {Fran{\c c}ois}, {Plez}, {Ludwig}, {Caffau}, {Molaro}, {Depagne},
  {Andersen}, {Barbuy}, {Beers}, {Nordstr{\"o}m}, \& {Primas}}]{Bonifaciodwarf}
{Bonifacio}, P., {Spite}, M., {Cayrel}, R., {et~al.} 2009, \aap, 501, 519

\bibitem[{{Cacciari} {et~al.}(1992){Cacciari}, {Clementini}, \&
  {Fernley}}]{CCF1992}
{Cacciari}, C., {Clementini}, G., \& {Fernley}, J.~A. 1992, \apj, 396, 219

\bibitem[{{Caffau} {et~al.}(2005){Caffau}, {Bonifacio}, {Faraggiana}, {Fran{\c
  c}ois}, {Gratton}, \& {Barbieri}}]{cafsvovl}
{Caffau}, E., {Bonifacio}, P., {Faraggiana}, R., {et~al.} 2005, \aap, 441, 533

\bibitem[{{Cassisi} {et~al.}(2004){Cassisi}, {Castellani}, {Caputo}, \&
  {Castellani}}]{cassisi}
{Cassisi}, S., {Castellani}, M., {Caputo}, F., \& {Castellani}, V. 2004, \aap,
  426, 641

\bibitem[{{Cayrel}(1988)}]{cayrel1988}
{Cayrel}, R. 1988, in IAU Symposium, Vol. 132, The Impact of Very High S/N
  Spectroscopy on Stellar Physics, ed. G.~{Cayrel de Strobel} \& M.~{Spite},
  345

\bibitem[{{Cayrel} {et~al.}(2004){Cayrel}, {Depagne}, {Spite}, {Hill}, {Spite},
  {Fran{\c c}ois}, {Plez}, {Beers}, {Primas}, {Andersen}, {Barbuy},
  {Bonifacio}, {Molaro}, \& {Nordstr{\"o}m}}]{firstV}
{Cayrel}, R., {Depagne}, E., {Spite}, M., {et~al.} 2004, \aap, 416, 1117

\bibitem[{{Collet} {et~al.}(2006){Collet}, {Asplund}, \& {Trampedach}}]{collet}
{Collet}, R., {Asplund}, M., \& {Trampedach}, R. 2006, {3D Hydrodynamical
  Simulations of Convection in Red-Giants Stellar Atmospheres}, ed. {Randich,
  S.~\& Pasquini, L.}, 306--+

\bibitem[{{Depagne} {et~al.}(2002){Depagne}, {Hill}, {Spite}, {Spite}, {Plez},
  {Beers}, {Barbuy}, {Cayrel}, {Andersen}, {Bonifacio}, {Fran{\c c}ois},
  {Nordstr{\"o}m}, \& {Primas}}]{firstII}
{Depagne}, E., {Hill}, V., {Spite}, M., {et~al.} 2002, \aap, 390, 187

\bibitem[{{Drawin}(1969)}]{drawin}
{Drawin}, H.~W. 1969, Zeitschrift fur Physik, 228, 99

\bibitem[{{Fabbian} {et~al.}(2009){Fabbian}, {Asplund}, {Barklem}, {Carlsson},
  \& {Kiselman}}]{fabbian}
{Fabbian}, D., {Asplund}, M., {Barklem}, P.~S., {Carlsson}, M., \& {Kiselman},
  D. 2009, \aap, 500, 1221

\bibitem[{{For} \& {Sneden}(2010)}]{FS10}
{For}, B. \& {Sneden}, C. 2010, ArXiv e-prints

\bibitem[{{Fran{\c c}ois} {et~al.}(2007){Fran{\c c}ois}, {Depagne}, {Hill},
  {Spite}, {Plez}, {Beers}, {James}, {Barbuy}, {Cayrel}, {Andersen},
  {Bonifacio}, {Molaro}, {Nordstr{\"o}m}, \& {Primas}}]{firstVIII}
{Fran{\c c}ois}, P., {Depagne}, E., {Hill}, V., {et~al.} 2007, \aap, 476, 935

\bibitem[{{Fran{\c c}ois} {et~al.}(2003){Fran{\c c}ois}, {Depagne}, {Hill},
  {Spite}, {Spite}, {Plez}, {Beers}, {Barbuy}, {Cayrel}, {Andersen},
  {Bonifacio}, {Molaro}, {Nordstr{\"o}m}, \& {Primas}}]{firstIII}
{Fran{\c c}ois}, P., {Depagne}, E., {Hill}, V., {et~al.} 2003, \aap, 403, 1105

\bibitem[{{Grevesse} \& {Sauval}(1998)}]{GrevSauv}
{Grevesse}, N. \& {Sauval}, A.~J. 1998, Space Science Reviews, 85, 161

\bibitem[{{Gustafsson} {et~al.}(2008){Gustafsson}, {Edvardsson}, {Eriksson},
  {J{\o}rgensen}, {Nordlund}, \& {Plez}}]{Gus08}
{Gustafsson}, B., {Edvardsson}, B., {Eriksson}, K., {et~al.} 2008, \aap, 486,
  951

\bibitem[{{Heger} \& {Woosley}(2002{\natexlab{a}})}]{heger+woosley}
{Heger}, A. \& {Woosley}, S.~E. 2002{\natexlab{a}}, in Nuclear Astrophysics,
  ed. W.~{Hillebrandt} \& E.~{M{\"u}ller}, 8--13

\bibitem[{{Heger} \& {Woosley}(2002{\natexlab{b}})}]{heger02}
{Heger}, A. \& {Woosley}, S.~E. 2002{\natexlab{b}}, \apj, 567, 532

\bibitem[{{Hill} {et~al.}(2002){Hill}, {Plez}, {Cayrel}, {Beers},
  {Nordstr{\"o}m}, {Andersen}, {Spite}, {Spite}, {Barbuy}, {Bonifacio},
  {Depagne}, {Fran{\c c}ois}, \& {Primas}}]{firstI}
{Hill}, V., {Plez}, B., {Cayrel}, R., {et~al.} 2002, \aap, 387, 560

\bibitem[{{Izutani} {et~al.}(2009){Izutani}, {Umeda}, \& {Tominaga}}]{Izu}
{Izutani}, N., {Umeda}, H., \& {Tominaga}, N. 2009, \apj, 692, 1517

\bibitem[{{Jorgensen} {et~al.}(1996){Jorgensen}, {Larsson}, {Iwamae}, \&
  {Yu}}]{JLI96}
{Jorgensen}, U.~G., {Larsson}, M., {Iwamae}, A., \& {Yu}, B. 1996, \aap, 315,
  204

\bibitem[{{Kobayashi} {et~al.}(2006){Kobayashi}, {Umeda}, {Nomoto}, {Tominaga},
  \& {Ohkubo}}]{galchem}
{Kobayashi}, C., {Umeda}, H., {Nomoto}, K., {Tominaga}, N., \& {Ohkubo}, T.
  2006, \apj, 653, 1145

\bibitem[{{Kolenberg} {et~al.}(2010){Kolenberg}, {Fossati}, {Shulyak},
  {Pikall}, {Barnes}, {Kochukhov}, \& {Tsymbal}}]{kolenberg}
{Kolenberg}, K., {Fossati}, L., {Shulyak}, D., {et~al.} 2010, \aap, 519, A64+

\bibitem[{{Lai} {et~al.}(2008){Lai}, {Bolte}, {Johnson}, {Lucatello}, {Heger},
  \& {Woosley}}]{lai08}
{Lai}, D.~K., {Bolte}, M., {Johnson}, J.~A., {et~al.} 2008, \apj, 681, 1524

\bibitem[{{Lambert} {et~al.}(1996){Lambert}, {Heath}, {Lemke}, \&
  {Drake}}]{LHL96}
{Lambert}, D.~L., {Heath}, J.~E., {Lemke}, M., \& {Drake}, J. 1996, \apjs, 103,
  183

\bibitem[{{Limongi} {et~al.}(2000){Limongi}, {Straniero}, \&
  {Chieffi}}]{limongi}
{Limongi}, M., {Straniero}, O., \& {Chieffi}, A. 2000, \apjs, 129, 625

\bibitem[{{Luque}(1996)}]{LC99}
{Luque}, J., C. D.~R. 1996, SRI Report, MP 96-001,

\bibitem[{{Mashonkina} \& {Gehren}(2001)}]{mash01}
{Mashonkina}, L. \& {Gehren}, T. 2001, \aap, 376, 232

\bibitem[{{Mashonkina} {et~al.}(2010){Mashonkina}, {Gehren}, {Shi}, {Korn}, \&
  {Grupp}}]{Mashonkina10}
{Mashonkina}, L., {Gehren}, T., {Shi}, J., {Korn}, A., \& {Grupp}, F. 2010, in
  IAU Symposium, Vol. 265, IAU Symposium, ed. {K.~Cunha, M.~Spite, \&
  B.~Barbuy}, 197--200

\bibitem[{{Mashonkina} {et~al.}(2007){Mashonkina}, {Korn}, \&
  {Przybilla}}]{Mashon07}
{Mashonkina}, L., {Korn}, A.~J., \& {Przybilla}, N. 2007, \aap, 461, 261

\bibitem[{{Matteucci}(2007)}]{matteucci}
{Matteucci}, F. 2007, in Astronomical Society of the Pacific Conference Series,
  Vol. 374, From Stars to Galaxies: Building the Pieces to Build Up the
  Universe, ed. {A.~Vallenari, R.~Tantalo, L.~Portinari, \& A.~Moretti}, 89--+

\bibitem[{{McWilliam} {et~al.}(1995){McWilliam}, {Preston}, {Sneden}, \&
  {Shectman}}]{McWilliam}
{McWilliam}, A., {Preston}, G.~W., {Sneden}, C., \& {Shectman}, S. 1995, \aj,
  109, 2736

\bibitem[{{Nemec}(2004)}]{Nem04}
{Nemec}, J.~M. 2004, \aj, 127, 2185

\bibitem[{{Nissen} {et~al.}(2007){Nissen}, {Akerman}, {Asplund}, {Fabbian},
  {Kerber}, {Kaufl}, \& {Pettini}}]{svovlnis}
{Nissen}, P.~E., {Akerman}, C., {Asplund}, M., {et~al.} 2007, \aap, 469, 319

\bibitem[{{Nomoto} {et~al.}(2005){Nomoto}, {Tominaga}, {Umeda}, \&
  {Kobayashi}}]{nomotoSNHN}
{Nomoto}, K., {Tominaga}, N., {Umeda}, H., \& {Kobayashi}, C. 2005, in IAU
  Symposium, Vol. 228, From Lithium to Uranium: Elemental Tracers of Early
  Cosmic Evolution, ed. V.~{Hill}, P.~{Fran{\c c}ois}, \& F.~{Primas}, 287--296

\bibitem[{{Paunzen} {et~al.}(1999){Paunzen}, {Andrievsky}, {Chernyshova},
  {Klochkova}, {Panchuk}, \& {Handler}}]{paunzen}
{Paunzen}, E., {Andrievsky}, S.~M., {Chernyshova}, I.~V., {et~al.} 1999, \aap,
  351, 981

\bibitem[{{Pe{\~n}a} {et~al.}(2009){Pe{\~n}a}, {Arellano Ferro}, {Pe{\~n}a
  Miller}, {Sareyan}, \& {{\'A}lvarez}}]{PAP09}
{Pe{\~n}a}, J.~H., {Arellano Ferro}, A., {Pe{\~n}a Miller}, R., {Sareyan},
  J.~P., \& {{\'A}lvarez}, M. 2009, \rmxaa, 45, 191

\bibitem[{{Preston} {et~al.}(2006){Preston}, {Sneden}, {Thompson}, {Shectman},
  \& {Burley}}]{preston}
{Preston}, G.~W., {Sneden}, C., {Thompson}, I.~B., {Shectman}, S.~A., \&
  {Burley}, G.~S. 2006, \aj, 132, 85

\bibitem[{{Primas} {et~al.}(1994){Primas}, {Molaro}, \& {Castelli}}]{prim94}
{Primas}, F., {Molaro}, P., \& {Castelli}, F. 1994, \aap, 290, 885

\bibitem[{{Sandstrom} {et~al.}(2001){Sandstrom}, {Pilachowski}, \&
  {Saha}}]{SPS01}
{Sandstrom}, K., {Pilachowski}, C.~A., \& {Saha}, A. 2001, \aj, 122, 3212

\bibitem[{{Shi} {et~al.}(2009){Shi}, {Gehren}, {Mashonkina}, \& {Zhao}}]{Shi09}
{Shi}, J.~R., {Gehren}, T., {Mashonkina}, L., \& {Zhao}, G. 2009, \aap, 503,
  533

\bibitem[{{Sivarani} {et~al.}(2004){Sivarani}, {Bonifacio}, {Molaro}, {Cayrel},
  {Spite}, {Spite}, {Plez}, {Andersen}, {Barbuy}, {Beers}, {Depagne}, {Hill},
  {Fran{\c c}ois}, {Nordstr{\"o}m}, \& {Primas}}]{firstIV}
{Sivarani}, T., {Bonifacio}, P., {Molaro}, P., {et~al.} 2004, \aap, 413, 1073

\bibitem[{{Smith}(1995)}]{Smi95}
{Smith}, H.~A. 1995, Cambridge Astrophysics Series, 27

\bibitem[{{Spite} {et~al.}(2006){Spite}, {Cayrel}, {Hill}, {Spite}, {Fran{\c
  c}ois}, {Plez}, {Bonifacio}, {Molaro}, {Depagne}, {Andersen}, {Barbuy},
  {Beers}, {Nordstr{\"o}m}, \& {Primas}}]{first9}
{Spite}, M., {Cayrel}, R., {Hill}, V., {et~al.} 2006, \aap, 455, 291

\bibitem[{{Spite} {et~al.}(2005){Spite}, {Cayrel}, {Plez}, {Hill}, {Spite},
  {Depagne}, {Fran{\c c}ois}, {Bonifacio}, {Barbuy}, {Beers}, {Andersen},
  {Molaro}, {Nordstr{\"o}m}, \& {Primas}}]{firstVI}
{Spite}, M., {Cayrel}, R., {Plez}, B., {et~al.} 2005, \aap, 430, 655

\bibitem[{{Spite} \& {Spite}(1978)}]{oldspite}
{Spite}, M. \& {Spite}, F. 1978, \aap, 67, 23

\bibitem[{{Steenbock} \& {Holweger}(1984)}]{steen84}
{Steenbock}, W. \& {Holweger}, H. 1984, \aap, 130, 319

\bibitem[{{Takeda} {et~al.}(2005){Takeda}, {Hashimoto}, {Taguchi}, {Yoshioka},
  {Takada-Hidai}, {Saito}, \& {Honda}}]{SNLTEtakeda}
{Takeda}, Y., {Hashimoto}, O., {Taguchi}, H., {et~al.} 2005, \pasj, 57, 751

\bibitem[{{Tominaga} {et~al.}(2007){Tominaga}, {Umeda}, \& {Nomoto}}]{SN13-50}
{Tominaga}, N., {Umeda}, H., \& {Nomoto}, K. 2007, \apj, 660, 516

\bibitem[{{Truran}(1988)}]{Tru}
{Truran}, J.~W. 1988, in IAU Symposium, Vol. 132, The Impact of Very High S/N
  Spectroscopy on Stellar Physics, ed. {G.~Cayrel de Strobel \& M.~Spite},
  577--+

\bibitem[{{Wanajo} {et~al.}(2010){Wanajo}, {Janka}, \& {Mueller}}]{wan2010}
{Wanajo}, S., {Janka}, H., \& {Mueller}, B. 2010, ArXiv e-prints

\bibitem[{{Woosley} \& {Weaver}(1995)}]{WW95}
{Woosley}, S.~E. \& {Weaver}, T.~A. 1995, \apjs, 101, 181

\bibitem[{{Zhang} {et~al.}(2008){Zhang}, {Gehren}, \& {Zhao}}]{ZhangSc}
{Zhang}, H.~W., {Gehren}, T., \& {Zhao}, G. 2008, \aap, 481, 489

\end{thebibliography}

\begin{thebibliography}{}
\bibitem[Bard et al.(1991)]{1991A&A...248..315B} Bard, A., Kock, A., \& 
Kock, M.\ 1991, \aap, 248, 315 (BKK)
\bibitem[Bridges(1973)]{1973pig..conf..418B} Bridges, J.~M.\ 1973, 
Phenomena in Ionized Gases, Eleventh International Conference, 418 (B)
\bibitem[Fuhr, Martin \& Wiese]{FMW}  Fuhr, J.R., Martin, G.A., and Wiese, W.L. \ 1988
Journal of Physical and Chemical Reference Data, 17, Suppl. 4 (FMW)
\bibitem[O'Brian et al.(1991)]{1991JOSAB...8.1185O} O'Brian, T.~R., 
Wickliffe, M.~E., Lawler, J.~E., Whaling, W., \& Brault, J.~W.\ 1991, 
Journal of the Optical Society of America B: Optical Physics, Volume 8, 
Issue 6, June 1991, pp.1185-1201, 8, 1185 (BWL)
\end{thebibliography}

\Online
\appendix

{\scriptsize
\begin{longtable}{llllllll}
\caption{\label{fe_1} Equivalent widths and abundances of iron lines}\\
\hline\hline\\
          &        &        &      & 
\multispan{2}{CS~30317-056\hfill}&\multispan{2}{CS~22881-039\hfill}\\
$\lambda$& $\chi$ & log gf & Ref. & EW  & A(Fe) & EW  & A(Fe)  \\
 nm      & eV     &        &      & pm  & dex   & pm  & dex    \\
\hline\hline\\
\endfirsthead \caption{continued.}\\ \hline\hline\\
          &        &        &      & 
\multispan{2}{CS~30317-056\hfill}&\multispan{2}{CS~22881-039\hfill}\\
$\lambda$& $\chi$ & log gf & Ref. & EW  & A(Fe) & EW  & A(Fe)  \\
 nm      & eV     &        &      & pm  & dex   & pm  & dex    \\
\hline\hline\\ \endhead \hline\hline \endfoot 
\ion{Fe}{i}\\
 337.0783  & 2.69 &$-0.266$  & BWL & \nod   & \nod   & \nod & \nod    \\                   
 339.9333  & 2.20 &$ -0.622 $& BWL &   1.54 &  3.57  & \nod & \nod    \\                   
 340.1519  & 0.92 &$ -2.059 $& BWL &   1.37 &  3.54  & \nod & \nod    \\                   
 340.7460  & 2.18 &$ -0.020 $& BWL &   3.68 &  3.52  & 1.76 &  3.96   \\                   
 341.3132  & 2.20 &$ -0.404 $& BWL &   2.03 &  3.51  & \nod & \nod    \\                   
 341.7841  & 2.22 &$ -0.676 $& BWL &   \nod &  \nod  &\nod  &\nod     \\                   
 341.8507  & 2.22 &$ -0.761 $& BWL &   1.12 &  3.57  &\nod  &\nod     \\                   
 342.4284  & 2.18 &$ -0.703 $& BWL &   \nod &  \nod  & \nod & \nod    \\                   
 342.5010  & 3.05 &$ -0.500 $& BWL &   \nod &  \nod  &\nod  &\nod     \\                   
 342.6383  & 0.99 &$ -1.909 $& BWL &   \nod &  \nod  &\nod  &\nod     \\                   
 342.7119  & 2.18 &$ -0.098 $& BWL &   3.48 &  3.54  &\nod  &\nod     \\                   
 342.8193  & 2.20 &$ -0.822 $& BWL &   0.82 &  3.44  &\nod  &\nod     \\                   
 344.0606  & 0.00 &$ -0.673 $& BWL &  10.11 &  3.60  & 7.22 &  4.08   \\                   
 344.0989  & 0.05 &$ -0.958 $& BWL &   8.83 &  3.64  & 6.00 &  4.03   \\                   
 344.3876  & 0.09 &$ -1.374 $& BWL &   7.93 &  3.83  & 3.79 &  3.90   \\                   
 344.5149  & 2.20 &$ -0.535 $& BWL &   2.22 &  3.69  & \nod & \nod    \\                   
 345.0328  & 2.22 &$ 0.902  $& BWL &   \nod &  \nod  & \nod & \nod    \\                   
 345.2275  & 0.96 &$ -1.919 $& BWL &   1.65 &  3.54  & \nod & \nod    \\                   
 347.5450  & 0.09 &$ -1.054 $& BWL &   8.53 &  3.67  & 4.94 &  3.85   \\                   
 347.6702  & 0.12 &$ -1.507 $& BWL &   7.22 &  3.76  & 4.07 &  4.12   \\                   
 348.5340  & 2.20 &$ -1.149 $& BWL & \nod   & \nod   & \nod & \nod    \\                   
 349.0574  & 0.05 &$ -1.105 $& BWL &   9.11 &  3.84  & 4.96 &  3.87   \\                   
 349.7841  & 0.11 &$ -1.549 $& BWL &   7.07 &  3.73  & 2.62 &  3.81   \\                   
 352.1261  & 0.92 &$ -0.988 $& BWL &   6.16 &  3.74  & 2.15 &  3.87   \\                   
 353.3198  & 2.88 &$ -0.112 $& BWL & \nod   & \nod   & \nod & \nod    \\                   
 353.6556  & 2.88 &$ +0.115 $& BWL &   1.71 &  3.60  & \nod & \nod    \\                   
 354.1083  & 2.85 &$ +0.252 $& BWL &   \nod &  \nod  & \nod & \nod    \\                   
 354.2076  & 2.87 &$ +0.207 $& BWL &   2.09 &  3.61  & \nod & \nod    \\                   
 355.3739  & 3.57 &$ +0.269 $& BWL &   1.07 &  3.95  & \nod & \nod    \\                   
 355.4118  & 0.96 &$ -2.206 $& BWL &   1.80 &  3.86  & \nod & \nod    \\                   
 355.4925  & 2.83 &$ +0.538 $& BWL &   2.75 &  3.42  & \nod & \nod    \\                   
 355.6878  & 2.85 &$ -0.040 $& FMW &   2.18 &  3.87  & \nod & \nod    \\                   
 356.5379  & 0.96 &$ -0.133 $& BWL &   7.86 &  3.43  & 4.90 &  3.72   \\                   
 358.1193  & 0.86 &$ +0.406 $& FMW &  11.12 &  3.49  & 7.43 &  3.81   \\                   
 358.4659  & 2.69 &$ -0.157 $& BWL &   1.57 &  3.62  & \nod & \nod    \\                   
 358.5319  & 0.96 &$ -0.802 $& BWL &   6.79 &  3.77  & 2.60 & 3.84    \\                   
 358.5705  & 0.92 &$ -1.187 $& FMW &   5.45 &  3.72  & \nod & \nod    \\                   
 358.6113  & 3.24 &$ +0.173 $& BWL &   1.18 &  3.73  & \nod & \nod    \\                   
 358.6985  & 0.99 &$ -0.796 $& BWL &   \nod &  \nod  & \nod & \nod    \\                   
 358.9105  & 0.86 &$ -2.115 $& FMW &   1.96 &  3.71  & \nod & \nod    \\                   
 360.3204  & 2.69 &$ -0.256 $& BWL &   1.11 &  3.54  & \nod & \nod    \\                   
 360.6679  & 2.69 &$ +0.323 $& BWL &   3.55 &  3.66  & \nod & \nod    \\                   
 360.8859  & 1.01 &$ -0.100 $& FMW &   8.36 &  3.57  & 4.61 & 3.66    \\                   
 361.0159  & 2.81 &$ +0.176 $& BWL &   2.06 &  3.57  & \nod & \nod    \\                   
 361.7786  & 3.02 &$ -0.029 $& BWL+BK& 1.39 &  3.78  & \nod & \nod    \\                   
 361.8768  & 0.99 &$ -0.003 $& BWL &   8.85 &  3.58  & 5.35 & 3.73    \\                   
 362.2003  & 2.76 &$ -0.150 $& BWL &   2.09 &  3.85  & \nod & \nod    \\                   
 362.3186  & 2.40 &$ -0.767 $& BWL &   1.24 &  3.80  & \nod & \nod    \\                   
 363.8296  & 2.76 &$ -0.375 $& BWL &    \nod&  \nod  & \nod & \nod    \\                   
 364.0389  & 2.73 &$ -0.107 $& BWL &   1.46 &  3.57  & \nod & \nod    \\                   
 364.7843  & 0.92 &$ -0.194 $& FMW &   8.45 &  3.52  & 5.66 & 3.76    \\                   
 380.5343  & 3.30 &$  0.312 $& BWL &   \nod & \nod   & \nod & \nod    \\                   
 380.6696  & 3.27 &$ +0.017 $& BWL & \nod   & \nod   & \nod & \nod    \\                   
 380.7537  & 2.22 &$ -0.992 $& BWL & \nod   & \nod   & \nod & \nod    \\                   
 381.5840  & 1.49 &$ +0.237 $& BWL &   8.56 &  3.63  & 5.67 &  3.82   \\                   
 381.6340  & 2.20 &$ -1.196 $& BWL &   0.68 &  3.65  & \nod & \nod    \\                   
 382.0425  & 0.86 &$ +0.119 $& FMW &  10.82 &  3.58  & 7.92 &  3.94   \\                   
 382.1178  & 3.27 &$ +0.198 $& BWL &   1.14 &  3.66  & \nod & \nod    \\                   
 382.5881  & 0.92 &$ -0.037 $& FMW &   9.89 &  3.62  & \nod & \nod    \\                   
 382.7823  & 1.56 &$ +0.062 $& FMW &   7.49 &  3.60  & \nod & \nod    \\                   
 384.0438  & 0.99 &$ -0.506 $& FMW &   \nod & \nod   & \nod & \nod    \\                   
 384.1048  & 1.61 &$ -0.045 $& BWL & \nod   & \nod   & \nod & \nod    \\                   
 384.3257  & 3.05 &$ -0.241 $& BWL & \nod   & \nod   & \nod & \nod    \\                   
 384.9967  & 1.01 &$ -0.871 $& FMW &   6.64 &  3.71  & 3.34 &  3.99   \\                   
 385.0818  & 0.99 &$ -1.734 $& FMW &   3.21 &  3.72  & 0.55 &  3.87   \\                   
 385.2573  & 2.18 &$ -1.185 $& BWL &    .70 &  3.63  & \nod & \nod    \\                   
 385.6372  & 0.05 &$ -1.286 $& FMW &   9.26 &  3.85  & 5.45 &  3.96   \\                   
 385.9213  & 2.40 &$ -0.749 $& BWL &   1.10 &  3.66  & \nod & \nod    \\                   
 385.9911  & 0.00 &$ -0.710 $& FMW &  11.39 &  3.66  & 8.22 & 4.07    \\                   
 386.5523  & 1.01 &$ -0.982 $& FMW &   6.30 &  3.72  & 2.91 & 4.01    \\                   
 386.7216  & 3.02 &$ -0.451 $& BWL &   0.73 &  3.82  & \nod & \nod    \\                   
 387.2501  & 0.99 &$ -0.928 $& FMW &   6.39 &  3.67  & 3.08 & 3.97    \\                   
 387.3761  & 2.43 &$ -0.876 $& BWL &   \nod & \nod   & \nod & \nod    \\                   
 387.8018  & 0.96 &$ -0.914 $& FMW &   6.90 &  3.76  & \nod & \nod    \\                   
 388.6282  & 0.05 &$ -1.076 $& FMW & \nod   & \nod   & \nod & \nod    \\                   
 388.7048  & 0.91 &$ -1.144 $& FMW & \nod   & \nod   & \nod & \nod    \\                   
 389.5656  & 0.11 &$ -1.670 $& FMW &   8.06 &  3.94  & \nod & \nod    \\                   
 389.9707  & 0.09 &$ -1.531 $& FMW &   8.47 &  3.90  & \nod & \nod    \\                   
 390.2946  & 1.56 &$ -0.466 $& FMW &   6.01 &  3.71  & 3.49 &  4.11   \\                   
 390.6480  & 0.11 &$ -2.243 $& FMW &   6.00 &  3.91  & 1.41 &  4.01   \\                   
 391.7181  & 0.99 &$ -2.155 $& FMW &   1.76 &  3.76  & \nod & \nod    \\                   
 392.0258  & 0.12 &$ -1.746 $& FMW &   7.62 &  3.89  & 3.27 & 4.02    \\                   
 392.7920  & 0.11 &$ -1.522 $& BWL &   8.70 &  3.97  & 4.32 & 4.00    \\                   
 394.0878  & 0.96 &$ -2.600 $& FMW &   1.14 &  3.94  & \nod & \nod    \\                   
 394.9953  & 2.18 &$ -1.251 $& BWL &   0.96 &  3.84  & \nod & \nod    \\                   
 395.6677  & 2.69 &$ -0.429 $& BWL &   1.31 &  3.73  & \nod & \nod    \\                   
 397.7741  & 2.20 &$ -1.119 $& BWL &   0.96 &  3.72  & \nod & \nod    \\                   
 399.7392  & 2.73 &$ -0.479 $& BWL &   1.03 &  3.69  & 0.72 &  4.32   \\                   
 400.5242  & 1.56 &$ -0.610 $& FMW &   5.10 &  3.61  & 2.27 &  3.97   \\                   
 400.9713  & 2.22 &$ -1.252 $& BWL &   0.73 &  3.75  & \nod & \nod    \\                   
 401.4531  & 3.05 &$ -0.587 $& BWL &   0.65 &  3.92  & \nod & \nod    \\                   
 402.1867  & 2.76 &$ -0.729 $& BWL &   0.61 &  3.72  & \nod & \nod    \\                   
 404.5812  & 1.49 &$ +0.280 $& FMW &   8.88 &  3.61  & 6.84 &  4.02   \\                   
 406.3594  & 1.56 &$ +0.062 $& BWL &   7.92 &  3.65  & 5.49 &  3.98   \\                   
 407.1738  & 1.61 &$ -0.022 $& FMW &   7.36 &  3.64  & 4.80 &  3.97   \\                   
 413.2058  & 1.61 &$ -0.675 $& BWL &   4.88 &  3.66  & 1.81 &  3.94   \\                   
 413.4678  & 2.83 &$ -0.649 $& BWL &   0.63 &  3.72  & \nod & \nod    \\                   
 414.3415  & 3.05 &$ -0.204 $& BWL &   0.75 &  3.59  & \nod & \nod    \\                   
 414.3868  & 1.56 &$ -0.511 $& BWL &   5.63 &  3.62  & 2.59 &  3.93   \\                   
 418.1755  & 2.83 &$ -0.371 $& BWL &   1.43 &  3.84  & \nod & \nod    \\                   
 418.7039  & 2.45 &$ -0.548 $& FMW &   1.36 &  3.58  & \nod & \nod    \\                   
 418.7795  & 2.42 &$ -0.554 $& FMW &   1.56 &  3.63  & \nod & \nod    \\                   
 419.1431  & 2.47 &$ -0.666 $& BWL &   1.15 &  3.63  & \nod & \nod    \\                   
 419.8304  & 2.40 &$ -0.719 $& FMW &   1.63 &  3.79  & \nod & \nod    \\                   
 419.9095  & 3.05 &$ +0.155 $& BWL &   1.94 &  3.71  & \nod & \nod    \\                   
 420.2029  & 1.49 &$ -0.708 $& FMW &   5.68 &  3.74  & \nod & \nod    \\                   
 421.0344  & 2.48 &$ -0.928 $& BWL &   0.82 &  3.74  & \nod & \nod    \\                   
 421.6184  & 0.00 &$ -3.356 $& FMW &   1.80 &  3.84  & \nod & \nod    \\                   
 421.9360  & 3.57 &$ +0.000 $& BWL &   0.55 &  3.80  & \nod & \nod    \\                   
 422.2213  & 2.45 &$ -0.967 $& FMW &   1.18 &  3.92  & \nod & \nod    \\                   
 422.7427  & 3.33 &$ +0.266 $& BWL &   1.08 &  3.60  & \nod & \nod    \\                   
 423.3603  & 2.48 &$ -0.604 $& FMW &   1.44 &  3.69  & \nod & \nod    \\                   
 423.5937  & 2.42 &$ -0.341 $& FMW &   2.21 &  3.60  & \nod & \nod    \\                   
 425.0119  & 2.47 &$ -0.405 $& FMW &   1.98 &  3.65  & \nod & \nod    \\                   
 425.0787  & 1.56 &$ -0.714 $& BWL &   5.02 &  3.66  & \nod & \nod    \\                   
 426.0474  & 2.40 &$ +0.109 $& BWL+BK& 4.18 &  3.56  & \nod & \nod    \\                   
 427.1154  & 2.45 &$ -0.349 $& FMW &   2.41 &  3.68  & \nod & \nod    \\                   
 427.1761  & 1.49 &$ -0.164 $& FMW &   7.87 &  3.74  & \nod & \nod    \\                   
 428.2403  & 2.18 &$ -0.779 $& BWL &   1.69 &  3.62  & \nod & \nod    \\                   
 429.9235  & 2.42 &$ -0.405 $& BWL &   \nod & \nod   & 0.25 & 3.46    \\                   
 432.5762  & 1.61 &$ +0.006 $& BWL &   7.77 &  3.67  & \nod & \nod    \\                   
 435.2735  & 2.22 &$ -1.287 $& BWL &   0.83 &  3.81  & \nod & \nod    \\                   
 437.5930  & 0.00 &$ -3.031 $& FMW &   3.17 &  3.85  & \nod & \nod    \\                   
 438.3545  & 1.49 &$ +0.200 $& FMW &   9.00 &  3.63  & \nod & \nod    \\                   
 440.4750  & 1.56 &$ -0.142 $& FMW &   7.32 &  3.63  & \nod & \nod    \\                   
 441.5123  & 1.61 &$ -0.615 $& FMW &   5.45 &  3.69  & \nod & \nod    \\                   
 442.7310  & 0.05 &$ -2.924 $& BWL &   2.65 &  3.67  & \nod & \nod    \\                   
 444.2339  & 2.20 &$ -1.255 $& FMW &   1.09 &  3.88  & \nod & \nod    \\                   
 445.9118  & 2.18 &$ -1.279 $& FMW &   0.98 &  3.82  & \nod & \nod    \\                   
 446.1653  & 0.09 &$ -3.210 $& FMW &   1.80 &  3.77  & \nod & \nod    \\                   
 448.2170  & 0.11 &$ -3.501 $& FMW &   1.48 &  3.97  & \nod & \nod    \\                   
 449.4563  & 2.20 &$ -1.136 $& FMW &   0.99 &  3.71  & \nod & \nod    \\                   
 452.8614  & 2.18 &$ -0.822 $& FMW &   2.05 &  3.75  & \nod & \nod    \\                   
 487.1318  & 2.87 &$ -0.363 $& BWL &   0.97 &  3.62  & \nod & \nod    \\                   
 487.2138  & 2.88 &$ -0.567 $& BWL &   0.58 &  3.60  & \nod & \nod    \\                   
 489.0755  & 2.88 &$ -0.394 $& BWL &   0.90 &  3.63  & 0.49 &  4.13   \\                   
 489.1492  & 2.85 &$ -0.112 $& BWL &   1.58 &  3.60  & 0.47 &  3.81   \\                   
 491.8994  & 2.87 &$ -0.342 $& BWL &   1.14 &  3.68  & 0.43 &  4.01   \\                   
 492.0503  & 2.83 &$ +0.068 $& BWL &   2.63 &  3.69  & 1.11 &  4.01   \\                   
 495.7299  & 2.85 &$ -0.408 $& BWL &   1.08 &  3.70  & 0.60 &  4.21   \\                   
 495.7597  & 2.81 &$ +0.233 $& BWL &   3.03 &  3.58  & 1.28 &  3.90   \\                   
 500.6119  & 2.83 &$ -0.638 $& BWL+BK& 0.62 &  3.64  & \nod & \nod    \\                   
 501.2068  & 0.86 &$ -2.642 $& FMW &   1.18 &  3.79  & \nod & \nod    \\                   
 504.1756  & 1.49 &$ -2.203 $& BWL &   0.75 &  3.82  & \nod & \nod    \\                   
 505.1635  & 0.92 &$ -2.795 $& FMW &   0.68 &  3.73  & \nod & \nod    \\                   
 511.0413  & 0.00 &$ -3.760 $& FMW &   0.97 &  3.84  & \nod & \nod    \\                   
 513.9463  & 2.94 &$ -0.509 $& BWL &   0.83 &  3.76  & \nod & \nod    \\                   
 517.1596  & 1.49 &$ -1.793 $& FMW &   1.44 &  3.72  & \nod & \nod    \\                   
 519.1455  & 3.04 &$ -0.551 $& BWL &   0.37 &  3.53  & \nod & \nod    \\                   
 519.2344  & 3.00 &$ -0.421 $& BWL &   0.57 &  3.56  & \nod & \nod    \\                   
 519.4942  & 1.56 &$ -2.090 $& FMW &   0.85 &  3.83  & \nod & \nod    \\                   
 522.7190  & 1.56 &$ -1.228 $& BWL &   3.24 &  3.71  & 0.77 &  3.94   \\                   
 523.2940  & 2.94 &$ -0.058 $& BWL &   1.60 &  3.63  & 0.96 &  4.16   \\                   
 526.6555  & 3.00 &$ -0.386 $& BWL &   0.65 &  3.58  & 0.28 &  3.97   \\                   
 526.9537  & 0.86 &$ -1.321 $& FMW &   7.42 &  3.95  & 2.48 &  3.99   \\                   
 527.0356  & 1.61 &$ -1.339 $& BWL &   2.80 &  3.78  & 0.51 &  3.91   \\                   
 532.4179  & 3.21 &$ -0.103 $& BKK &   0.66 &  3.53  & \nod & \nod    \\                   
 532.8039  & 0.92 &$ -1.466 $& FMW &   6.08 &  3.83  & \nod & \nod    \\                   
 532.8532  & 1.56 &$ -1.850 $& BWL &   1.31 &  3.80  & 1.66 &  3.96   \\                   
 534.1024  & 1.61 &$ -1.953 $& BWL &   1.11 &  3.87  & \nod & \nod    \\                   
 537.1490  & 0.96 &$ -1.645 $& FMW &   5.09 &  3.84  & 0.90 & 3.87    \\                   
 539.7128  & 0.92 &$ -1.993 $& FMW &   3.41 &  3.79  & \nod & \nod    \\                   
 540.5775  & 0.99 &$ -1.844 $& FMW &   3.52 &  3.74  & \nod & \nod    \\                   
 542.9697  & 0.96 &$ -1.879 $& FMW &   3.81 &  3.80  & 0.54 &  3.86   \\                   
 543.4524  & 1.01 &$ -2.122 $& FMW &   2.44 &  3.80  & \nod & \nod    \\                   
 544.6917  & 0.99 &$ -1.914 $& BWL &   3.54 &  3.82  & \nod & \nod    \\                   
 545.5609  & 1.01 &$ -2.091 $& BWL &   2.39 &  3.76  & \nod & \nod    \\                   
 561.5644  & 3.33 &$ +0.050 $& BKK &   0.92 &  3.65  & \nod & \nod    \\                   
\ion{Fe}{ii}\\
 423.3172  & 2.58 &$ -1.900 $ & av  &   1.06 &  3.74  & \nod & \nod \\                   
 492.3927  & 2.89 &$ -1.320 $ & av  &   1.87 &  3.74  & 1.34 & 4.09 \\                   
 501.8440  & 2.89 &$ -1.220 $ & B   &   2.38 &  3.78  & \nod & \nod \\                   
 516.9033  & 2.89 &$ -0.870 $ & FMW &   3.34 &  3.64  & \nod & \nod \\                   
\hline\hline
\end{longtable}

\normalsize
\bibliographystyle{aa}
{}

\section{Line Lists}\label{linelist}

{\scriptsize
\begin{longtable}{llllllllllll}
\caption{The tabel includes the detected elements of CS 22881-039, compared to \citet{preston} and listed next to CS 30317-056. Wavelength, excitation potential ($\chi$), log gf and the obtained equaivalent width with associated abundance and uncertainty of each line is also given. ** means that the abundance was also
found in Midas. The oscillator strengths listed for Li and Sc are values taken from VALD, but fine structure and hyperfine structure calculated log gf's were actually applied to obtain the abundances listed below. A 'ul' implies that the abundances are only upper limits. The Mg abundances were found by line fitting applying a rotational broadning of 21
km/s - the values listed after semi-colon. The NLTE correction for O has been calculated by P. Bonifacio, while the corrections of Mg, have been calculated by S. Andrievsky. There are two different values of the Mg \dn, first mentioned are the corrections of CS 22881-039 and last mentioned of CS 30317-056.  All other estimates of the NLTE abundance corrections have been made based on literature (see Section \ref{sec:rareearth}) and a $\sim$ indicates that these corrections are slightly more uncertain. }\\
\hline\hline
 & & & & \multicolumn{3}{c}{CS~22881-039} & \multicolumn{1}{c}{  Preston  } & \multicolumn{3}{c}{  CS~30317-056  }\\
Element&$\lambda$& $\chi$ & log gf & EW    & Abun  & Error Abun &  EW & EW  & Abun & Error Abun & \dn \\
\hline\hline
\endfirsthead \caption{continued.}\\  
\hline\hline
 & & & & \multicolumn{3}{c}{CS~22881-039} &  \multicolumn{1}{c}{  Preston  } & \multicolumn{3}{c}{  CS~30317-056  }\\
Element&$\lambda$& $\chi$ & log gf  & EW &  Abun & Error Abun & EW & EW  & Abun & Error Abun & \dn \\
\hline\hline\\ \endhead \hline\hline \endfoot 
Li   & 610.37  & 1.85&  0.361&  1.0   &     0.6    &  ul   &    \nod &        &        &       &            \\
Li   & 670.79  & 0.00& -0.309&        &            &       &    \nod &  2.2   & 1.0    &  ul   &            \\
CH   & 430.80 & 0.16& -1.356&        &    6.50     &  ul   &    \nod &        & 6.50   &   ul	&            \\ 
O   1& 777.19 & 9.15&  0.369&        &            &       &    \nod &  16.8  & 6.88   &  0.24	&   -0.1     \\ 
Na  1& 589.0 & 0.00&  0.117&         &     ;3.46   &  0.13 &    \nod &  67.7  & 3.04   &  0.06 &   -0.2     \\
Na  1& 589.59 & 0.00& -0.184&        &     ;3.45   &  0.20 &    \nod &  31.4  & 2.85   &  0.07 &   -0.1      \\
Mg  1& 382.94 & 2.71& -0.207&  187.0 &  5.88 ;5.62&  0.04 &     140 & 141.6  & 5.06   &  0.02 &   0.02/0.13	\\
Mg  1& 383.23 & 2.71&  0.146&  196.0 &  5.66 ;5.62&  0.08 &    \nod & 171.2  & 5.27   &  0.19 &   -0.02/0.05   \\
Mg  1& 383.83 & 2.72&  0.415&  216.0 &  5.52 ;5.62&  0.03 &    \nod & 173.8  & 4.98   &  0.10	&   0.11/0.08   \\
Mg  1& 416.37 & 4.34& -1.000&        &             &       &    \nod& 17.70  & 5.45   & 0.20   &               \\
Mg  1& 435.19 & 4.34& -0.525&        &            &  	   &   \nod & 28.50  & 5.21   & 0.13 	&              \\
Mg  1& 457.11 & 0.00& -5.393&        &            &  	   &   \nod &  6.30  & 5.28   & 0.10 	&              \\
Mg  1& 470.30 &     &       &        &            &  	    &    33   &        &        &  	&              \\
Mg  1& 517.27 & 2.71& -0.380&  179.0 &  5.56 ;5.59&  0.11 &    \nod &        &        &  	&  -0.01/0.03 	\\
Mg  1& 518.36 & 2.72& -0.158&  200.0 &  5.72 ;5.59&  0.02 &     \nod&        &        &  	&  -0.17/-0.08 	\\
Mg  1& 552.84 & 4.34& -0.341&   23.0 &  4.87 ;5.21&  0.09 &     40  &        &        &  	&   0.11/0.11   \\
Al  1& 394.40 & 0.00& -0.640&   56.0 &    2.94    &  0.09 &    \nod &  62.2  & 3.06   &  0.05 &   0.7     \\
Al  1& 396.15 & 0.01& -0.340&   86.6 &    3.02    &  0.07 &      86 &  69.4  & 2.85   &  0.03 &   0.7      \\
Si  1& 390.55 & 1.91& -1.090&  122.3 & 4.81,4.6** &  0.05 &      111& 115.9  & 4.75   &  0.02 &   -0.05   	  \\
S   1& 921.29 & 6.53&  0.420&        &            &       &     \nod&  51.3  & 5.12   &  0.07 &   -0.4 	  \\ 
Ca  1& 422.67 & 0.00&  0.240&  192.0 &     4.43   &  0.08 &     167 & 151.4  & 3.76   &  0.07 &   0.02    \\
Ca  1& 428.30 & 1.89& -0.220&        &            &       &     \nod&  26.2  & 4.10   &  0.20 &		\\ 
Ca  1& 431.87 & 1.90& -0.210&   20.3 &     3.93   &  0.08 &     \nod&  17.4  & 3.89   &  0.05 &          \\
Ca  1& 442.54 & 1.88& -0.360&        &            &       &     \nod&  11.4  & 3.81   &  0.05 &           \\
Ca  1& 443.57 & 1.89& -0.520&        &            &       &     \nod&  12.5  & 4.02   &  0.07 &  	   \\ 
Ca  1& 445.48 & 1.90&  0.260&   44.2 &     3.89   &  0.06 &    \nod &  43.1  & 3.90   &  0.04 &       	\\
Ca  1& 558.88  & 2.52&  0.210&        &            &       &      16 &  18.2  & 4.01   &  0.06 &  0.27   \\
Ca  1& 616.22 & 1.90& -0.090&   21.9 &     3.78   &  0.07 &    \nod &        &        &       &  0.19   \\
Ca  1& 643.91  & 2.52&  0.470&   11.2 &     3.47   &  0.13 &    \nod &        &        &       &  0.22   \\	
Ca  1& 649.38  &     &       &        &            &       &     17  &        &        &       &         \\
Sc  2& 424.68 & 0.31&  0.240&  106.0 &    0.53    &  0.05 &     105 &  96.1  & 0.37   &  0.05	&	     \\
Sc  2& 431.41 & 0.62& -0.100&   49.9 &    0.47    &  0.03 &    \nod &  51.7  & 0.46   &  0.10	&	  \\
Sc  2& 440.04 & 0.61& -0.536&        &            &       &     28  &  25.3  & 0.46 	&  0.10	&	    \\
Sc  2& 441.56 & 0.60& -0.670&        &            &       &     \nod & 22.7  & 0.55     & 0.20 &          \\
Sc  2& 552.68  & 1.77& -0.030&        &            &       &       13&  13.5  & 0.79 	&  0.09	&	   \\ 
Ti  1& 395.82  & 0.05& -0.177&        &            &       &       20&  20.5  &  2.94  &  0.18 &       \\
Ti  1& 399.86 & 0.05& -0.060&        &            &       &     \nod&  12.5  &  2.58  &  0.08 &          \\ 
Ti  1& 453.32  & 0.84&  0.476&        &            &       &       13&  17.6  &  2.92  &  0.04 &           \\
Ti  1& 498.17  & 0.84&  0.50 &   12.0 &    2.65    &  0.21 &     \nod&  11.2  &  2.65  &  0.08 &          \\
Ti  1& 499.11 & 0.84&  0.380&        &            &       &    \nod &  11.6  &  2.78  &  0.06 &           \\ 
Ti  1& 499.95 & 0.83&  0.250&        &            &       &    \nod &  12.4  &  2.94  &  0.10 &           \\ 
Ti  2& 375.93 & 0.61&  0.270&  200.0 &    3.07    &  0.30 &     \nod&	189.6  &  2.97  &  0.08 &          \\
Ti  2& 376.13 & 0.57&  0.170&  200.8 &    3.14    &  0.29 &     \nod&	184.9  &  2.96  &  0.05 &          \\
Ti  2& 391.35 & 1.12& -0.410&  147.2 &    3.15    &  0.20 &     \nod& 128.6  &  2.90  &  0.04 &           \\
Ti  2& 401.24 & 0.57& -1.750&   53.0 &    2.50    &  0.13 &     \nod&	 57.3  &  2.70  &  0.02 &          \\
Ti  2& 402.83 & 1.89& -0.990&   23.3 &    2.53    &  0.14 &     \nod&  28.0  &  2.77  &  0.03 &           \\
Ti  2& 429.02 & 1.16& -0.930&   76.4 &    2.51    &  0.10 &     \nod&  83.7  &  2.74  &  0.06 &           \\
Ti  2& 430.00 & 1.18& -0.490&  113.9 &    2.59    &  0.08 &     \nod& 110.1  &  2.66  &  0.02 &          \\ 
Ti  2& 430.19  &     &       &        &            &       &      54 &  51.8  &  2.62  &  0.04 &          \\
Ti  2& 433.79 & 1.08& -0.980&   70.8 &    2.41    &  0.05 &     \nod&	 59.1  &  2.41  &  0.05 &           \\
Ti  2& 439.41  & 1.22& -1.770&        &            &       &      18 &  23.7  &  2.79  &  0.07 &           \\
Ti  2& 439.50 & 1.08& -0.510&  125.8 &    2.70    &  0.04 &     \nod& 117.6  &  2.69  &  0.03 &          \\
Ti  2& 439.59 & 1.24& -1.970&        &            &       &      14 &  18.7  &  2.89  &  0.03 &          \\ 
Ti  2& 439.98 & 1.24& -1.220&   46.8 &    2.49    &  0.04 &     \nod&  47.7  &  2.65  &  0.04 &          \\     
Ti  2& 441.77 & 1.16& -1.230&   41.9 &    2.36    &  0.03 &     \nod&	 60.3  &  2.75  &  0.03 &          \\ 
Ti  2& 444.38 & 1.08& -0.700&  102.0 &    2.51    &  0.14 &      101& 104.9  &  2.69  &  0.04 &           \\
Ti  2& 444.46 & 1.12& -2.210&        &            &       &    \nod &  11.8  &  2.79  &  0.06 &           \\ 
Ti  2& 445.05 & 1.08& -1.510&        &            &       &    \nod &  42.3  &  2.71  &  0.06 &        	\\ 
Ti  2& 446.45 & 1.16& -1.810&        &            &       &    \nod &  26.8  &  2.83  &  0.04 &        	\\ 
Ti  2& 446.85 & 1.13& -0.600&  109.3 &    2.57    &  0.07 &     106 & 108.6  &  2.68  &  0.04 &            \\
Ti  2& 450.13 & 1.12& -0.760&   91.3 &    2.46    &  0.06 &      99 &  96.5  &  2.66  &  0.06 &            \\
Ti  2& 453.40 & 1.24& -0.540&  107.0 &    2.57    &  0.07 &    \nod & 107.5  &  2.70  &  0.04 &        	\\ 
Ti  2& 456.38 & 1.22& -0.790&   86.5 &    2.52    &  0.05 &    \nod &  85.2  &  2.65  &  0.02 &            \\ 
Ti  2& 457.20 & 1.57& -0.230&        &            &       &    \nod & 103.6  &  2.66  &  0.03 &            \\ 
Ti  2& 512.92 & 1.89& -1.300&        &            &       &    \nod &  13.4  &  2.64  &  0.11 &            \\ 
Ti  2& 518.60 & 1.89& -1.370&        &            &       &    \nod &  14.6  &  2.75  &  0.03 &	     \\ 
Ti  2& 518.87 & 1.58& -1.050&   34.5 &    2.42    &  0.10 &    \nod &  34.8  &  2.57  &  0.03	&	     \\ 
Ti  2& 522.65 & 1.57& -1.230&        &            &       &    \nod &  25.4  &  2.57  &  0.04	&	     \\ 
Ti  2& 533.68 & 1.58& -1.630&        &            &       &    \nod &  15.2  &  2.72  &  0.08	&	      \\ 
Ti  2& 541.88  &     &       &        &            &       &      8  &        &        &       &             \\
V   2& 395.20  & 1.48& -0.784&        &            &       &      11 &        &        &       &	      \\
V   2& 400.57  & 1.82& -0.522&        &            &       &      11 &        &        &       &	      \\
Cr  1& 425.43 & 0.00& -0.110&   52.6 &      2.48  &  0.02 &       69&  56.6  &  2.59  &  0.10 &   0.4      \\
Cr  1& 427.48 & 0.00& -0.230&   47.6 &      2.53  &  0.06 &      52 &  55.0  &  2.68  &  0.10 &   0.4      \\
Cr  1& 428.97  & 0.00&  0.361&        &            &       &      38 &  43.8  &  2.66  &  0.13 &   0.4      \\
Cr  1& 520.60 & 0.94&  0.020&   24.0 &      2.73  &  0.11 &   \nod  &         &        &       &           \\
Cr  1& 520.84 & 0.94&  0.160&   27.2 &      2.66  &  0.04 &    \nod &         &       &        &           \\
Mn  1& 403.08 & 0.00& -0.480&   37.2 &      2.03  &  0.16 &      37 &  28.8  &  1.95  &  0.08 &   0.6       \\
Mn  1& 403.31 & 0.00& -0.620&   30.5 &      2.06  &  0.19 &      31 &  19.6  &  1.89  &  0.05 &   0.6        \\
Mn  1& 403.45  & 0.00& -0.811&        &            &       &      17 &  13.0  &  1.88  &  0.10 &   0.6       \\
Co  1& 384.55 & 0.92&  0.010&   25.9 &      2.41  &  0.21 &    \nod &  18.2  &  2.28  &  0.06 & $\sim$0.6      \\
Co  1& 399.53 & 0.92& -0.22 &        &            &       &    \nod &  12.9  &  2.33  &  0.06 & $\sim$0.6     \\  
Co  1& 412.13 & 0.92& -0.320&   11.4 &      2.22  &  0.27 &      13 &  12.6  &  2.40  &  0.06 & $\sim$0.6      \\
Ni  1& 380.71 & 0.42& -1.180&   46.0 &      3.26  &  0.14 &   \nod  &  42.5  &  3.27  &  0.07 &           \\
Ni  1& 385.83 & 0.42& -0.970&   70.3 &      3.36  &  0.06 &    \nod &  59.8  &  3.29  &  0.05 &           \\
Ni  1& 547.69 & 1.83& -0.890&        &            &       &    \nod &  12.4  &  3.54  &  0.08 &             \\ 
Sr  2& 407.77 & 0.00&  0.170&  159.1 &     0.329  & 0.13  &     141 & 148.1  &  0.17  &  0.10 &  0.6      \\
Sr  2& 421.55  & 0.00& -0.145&  133.5 &     0.111  & 0.16  &     125 & 126.9  &  0.02  &  0.08 &  0.6      \\
Y   2& 378.87  &     &       &        &            &       &      19 &        &        &       &           \\
Ba  2& 455.40  & 0.00&  0.17 &        &    \nod    &       &     37  &  42.4  & -1.04  & -0.14 &  0.15     \\
Ba  2& 493.41  & 0.00& -0.150&  23.8  &    -1.240  &  0.09 &     25  &        &        &       &           \\
\hline\hline
\end{longtable}
}

\end{document}